\documentclass[aps,pre,psfig,twocolumn,showpacs,superscriptaddress,reprint,floatfix]{revtex4-1}
\usepackage{amsmath}
\usepackage{mathtools}
\usepackage{amssymb}
\usepackage{color}
\usepackage{graphics}
\usepackage{epsfig}
\usepackage[normalem]{ulem} %to strike the words
\usepackage{cancel}
\usepackage[mathlines]{lineno}
\graphicspath{{Figures/}}
\usepackage{xtab,afterpage,longtable}
\usepackage{booktabs}   
\usepackage{ltablex}
\relax
\usepackage{float}

\begin{document}
%%%%%%%%%%%%%%%%%%%%%%%%%%%%%%%%%%%%%%%%%%%%%%%%%%%%%%%%%%%%%%%%%%%%%%%%%%%%%%%%%%%%%%%%%%%%%%%%%%%%%%%%%%%%%%%%%%%%%%%%%%%%%%%%%%%%%%%%%%%%%%%%%%%%%%%%%%%%%%%%%%%%%%%%%%%%%%%%%%%%%%%%%%%%%%%%%%%%%%%%%%%%%%
\title{Synchronization of Dust Acoustic Waves in a forced Korteweg-de Vries-Burgers model}
%\author{ABC}
\author{Ajaz Mir}
\email{ajaz.mir@iitjammu.ac.in}
\affiliation{Indian Institute of Technology Jammu, Jammu, J\&K,  181221, India}
\author{Sanat Tiwari}
\email{sanat.tiwari@iitjammu.ac.in}
\affiliation{Indian Institute of Technology Jammu, Jammu, J\&K,  181221, India}
\author{Abhijit Sen}
\affiliation{Institute for Plasma Research, Gandhinagar, Gujarat, 382428, India}
\author{Chris Crabtree}
\affiliation{Naval Research Laboratory, Washington, DC 20375, USA}
\author{Gurudas Ganguli}
\affiliation{Naval Research Laboratory, Washington, DC 20375, USA}
\author{John Goree}
\affiliation{Department of Physics and Astronomy, University of Iowa, Iowa City, IA 52242, USA}
% %\author{Bin Liu}
% %\affiliation{Department of Physics and Astronomy, University of Iowa, Iowa City, IA 52242, USA}
\date{\today}
%%%%%%%%%%%%%%%%%%%%%%%%%%%%%%%%%%%%%%%%%%%%%%%%%%%%%%%%%%%%%%%%%%%%%%%%%%%%%%%%%%%%%%%%%%%%%%%%%%%%%%%%%%%%%%%%%%%%%%%%%%%%%%%%%%%%%%%%%%%%%%%%%%%%%%%%%%%%%%%%%%%%%%%%%%%%%%%%%%%%%%%%%%%%%%%%%%%%%%%%%%%%%%
\begin{abstract}
The synchronization of dust acoustic waves to an external periodic source is studied in the framework of a driven Korteweg-de Vries-Burgers equation that takes into account the appropriate nonlinear and dispersive nature of low frequency waves in a dusty plasma medium. For a spatio-temporally varying source term the system is shown to demonstrate  harmonic (1:1) and super-harmonic (1:2) synchronized states. The existence domains of these states are delineated in the form of Arnold tongue diagrams in the parametric space of the forcing amplitude and forcing frequency and their resemblance to some past experimental results is discussed. 
\end{abstract}
%%%%%%%%%%%%%%%%%%%%%%%%%%%%%%%%%%%%%%%%%%%%%%%%%%%%%%%%%%%%%%%%%%%%%%%%%%%%%%%%%%%%%%%%%%%%%%%%%%%%%%%%%%%%%%%%%%%%%%%%%%%%%%%%%%%%%%%%%%%%%%%%%%%%%%%%%%%%%%%%%%%%%%%%%%%%%%%%%%%%%%%%%%%%%%%%%%%%%%%%%%%%%%
\maketitle
%%%%%%%%%%%%%%%%%%%%%%%%%%%%%%%%%%%%%%%%%%%%%%%%%%%%%%%%%%%%%%%%%%%%%%%%%%%%%%%%%%%%%%%%%%%%%%%%%%%%%%%%%%%%%%%%%%%%%%%%%%%%%%%%%%%%%%%%%%%%%%%%%%%%%%%%%%%%%%%%%%%%%%%%%%%%%%%%%%%%%%%%%%%%%%%%%%%%%%%%%%%%%%
\section{Introduction}
\label{intro}
%%%%%%%%%%%%%%%%%%%%%%%%%%%%%%%%%%%%%%%%%%%%%%%%%%%%%%%%%%%%%%%%%%%%%%%%%%%%%%%%%%%%%%%%%%%%%%%%%%%%%%%%%%%%%%%%%%%%%%%%%%%%%%%%%%%%%%%%%%%%%%%%%%%%%%%%%%%%%%%%%%%%%%%%%%%%%%%%%%%%%%%%%%%%%%%%%%%%%%%%%%%%%%
%
\paragraph*{}
The nonlinear phenomenon of frequency synchronization is ubiquitous in many physical, chemical, and biological systems and has been the subject of a large number of studies over the past several years~\cite{Pikovsky_CUP_2001,Strogatz_PP_2004,Balanov_SV_2009}. The simplest mathematical model describing this phenomenon consists of an ensemble of globally coupled nonlinear point oscillators that adjust their intrinsic frequencies to a common collective frequency as the coupling strength is increased~\cite{Acebron_RMP_2005,Kuramoto_PTPS_1984,Aronson_PDNP_1990,Kuramoto_DP_2003}. Such a nonlinear phenomenon can also be observed in a continuum medium (a fluid) where a self-excited oscillation or a wave can interact with a driving force and adjust its oscillation or wave frequency~\cite{Williams_POP_2010,Block_PRE_2001,Klinger_PLA_1993,Klinger_PRE_1995,Gyergyek_PPCF_1999,Nurujjaman_PRE_2010}. A plasma system with its wide variety of collective modes and complex nonlinear dynamics provides a rich and challenging medium for the exploration of synchronization phenomena. A number of past experimental studies have examined the driven response of a plasma to an external frequency source~\cite{Suranga_PRE_2012, Pilch_POP_2009, Klinger_PRE_1995, Williams_PRE_2014A}. There have also been a few studies devoted to an investigation of mutual synchronization between two plasma devices~\cite{Neeraj_POP_2015,Fukuyama_PRL_2006,Neeraj_PRE_2016}. 
\paragraph*{}
More recently, synchronization phenomena have been experimentally explored in dusty plasma devices where it is easy to visualize the low-frequency wave activity using fast video imaging. A dusty plasma is a four-component plasma of electrons, ions, neutral gas atoms, and micron-size particles of solid matter~\cite{Shukla_RMP_2009, Morfill_RMP_2009, Shukla_IOP_2001}. \textcolor{black}{It can be produced in a laboratory device like a glow discharge plasma, by introducing micron sized solid particles ~\cite{Thompson_POP_1997, Schwabe_PRL_2007, Barkan_POP_1995, Heinrich_PRL_2009}. These small solid particles (dust) get negatively charged by absorbing more electrons which have a higher mobility  than ions. Such a charged medium consisting of dust, ions and electrons, can sustain a variety of collective modes~\cite{Bandyopadhyay_PRL_2008, Deka_POP_2017, Piel_PPCF_2001, Shukla_IOP_2001}. The dust acoustic wave (DAW) or dust density wave (DDW) is one such well known low frequency compressional mode~\cite{Liu_POP_2018, Shukla_IOP_2001}. A DAW can be spontaneously excited due to the onset of an   ion-streaming instability.} \textcolor{black}{The DAW has a very low frequency (typically 10–100 Hz)~\cite{Suranga_PRE_2012, Thompson_POP_1997} due to the large mass of the dust particles and can consequently be visually observed; through its images and video recording~\cite{Thompson_IEEE_TPS_1999,Don_IEEE_TPS_2014,Feng_RSI_2007,Schwabe_PRL_2007}.}
\paragraph*{}
Synchronization of dust acoustic waves has been studied in an anodic plasma~\citep{Pilch_POP_2009}, radio-frequency (RF) and direct-current (DC) plasmas~\citep{Suranga_PRE_2012, Deka_PST_2020, Williams_IEEE_2018}. Pilch~\textit{et al.}~\citep{Pilch_POP_2009} reported the entrainment of DAWs through a driving modulation to the anode. Ruhunusiri~\textit{et al.}~\citep{Suranga_PRE_2012} reported observation of harmonic, super-harmonic, and sub-harmonic synchrony of self-excited cnoidal DAWs. This was achieved through the driven modulation of the streaming ions in the dust cloud. Their experiments showed parametric regions for the occurrence of such synchrony in the form of Arnold tongue diagrams in the state space of the driving frequency and driving amplitude. They also observed features like the branching of the tongues and the existence of an amplitude threshold for synchronization to occur. Williams~\textit{et al.}~\citep{Williams_IEEE_2018} compared DAW synchronization in RF and DC generated plasmas. Their results suggested that in a RF plasma, synchronization was restricted to a part of the dust cloud volume unlike the complete dust cloud synchrony in a DC discharge plasma. Deka \textit{et al.}~\cite{Deka_PST_2020} observed the synchronization of self-excited DDW, through the suppression mechanism, by modulating ion streaming using an external sinusoidal driver. Recently, Liu~\textit{et al.}~\citep{Liu_IEEE_2021} carried out experiments in the Plasma Kristall-4 (PK-4) device on board the International Space Station (ISS) under micro-gravity conditions and reported phase locking for harmonic synchronization.
 \paragraph*{}
 Theoretical efforts towards interpretation and physical understanding of these experimental results have so far been limited to providing qualitative comparisons with results obtained from very simple dynamical models. One of the commonly employed mathematical model is the periodically driven van der Pol oscillator~\cite{Van_PM_1927,Pikovsky_CUP_2001,Balanov_SV_2009},
 %%%%%%%%%%%%%%%%%%%%%%%%%%%%%%%%%%%%%%%%%%%%%%%%%%
 \begin{equation}
     \frac{d^2 x}{dt^2} + (c_1 - c_2 x^2)\frac{d x}{dt} + \omega_0^2 x = A_{dr} \cos(2 \pi f_{dr} t)
\label{fVdP_eqn}
 \end{equation}
 %%%%%%%%%%%%%%%%%%%%%%%%%%%%%%%%%%%%%%%%%%%%%%%%%%
 which describes the displacement $x$ of a harmonic oscillator with a natural frequency $\omega_0$, with terms for a nonlinear damping $c_2 x^2 dx/dt$, a source of energy for self-excitation $c_1 dx/dt$, 
 and a periodic driving source at a frequency $f_{dr}$. The van der Pol oscillator can exhibit synchronization not only at $f_{dr}/f_0 \approx 1$, which is called ``harmonic'' synchronization, but at ratios that are rational numbers. If  $f_{dr}/f_0 > 1$, the synchronization is said to be ``super-harmonic'', whereas if  $f_{dr}/f_0 < 1$ it is ``sub-harmonic''. 
 %Often, the van der Pol oscillator has been used as a reference for characterizing properties of wave synchronization in plasmas and other mediums that support the propagation of waves~\cite{Kovalev_IJBC_2017,Neeraj_POP_2015,Suranga_PRE_2012,Deka_PST_2020}. The dynamics of propagating waves, however, differ from that of point oscillators, suggesting that Eq.~\eqref{fVdP_eqn} is not the best representation of nonlinear waves in plasmas. Hence, there is a need to develop theoretical descriptions of nonlinear synchronization of waves, as compared to point oscillators. 
 Often, the van der Pol oscillator model has been used as a reference for characterizing synchronization phenomena in plasmas and other media that support the propagation of waves~\cite{Kovalev_IJBC_2017,Neeraj_POP_2015,Suranga_PRE_2012,Deka_PST_2020}. However, as a point oscillator  model its dynamics is restricted to nonlinear oscillations and it cannot correctly represent nonlinear waves. This is also evident from the fact that the van der Pol model  is an ordinary differential equation in time and therefore has no spatial dynamics that characterizes a propagating wave. In addition, for nonlinear ion acoustic or dust acoustic waves dispersion plays an important role in defining their propagation characteristics.  Hence there is a need to develop appropriate theoretical models to describe nonlinear synchronization of such dispersive waves. 
 \paragraph*{}
In this paper, we present such a model and use it to demonstrate synchronization of dust acoustic waves. The model is a generalization of the forced Korteweg-de Vries (fKdV) model that was developed by Sen~{\it et al.}~\cite{Sen_ASR_2015} for driven nonlinear acoustic waves and subsequently modified and extensively used to study nonlinear precursor solitons in dusty plasma experiments~\cite{Sanat_POP_2016,Surabhi_PRE_2016}. In the original fKdV model~\cite{Sen_ASR_2015} the driving term was a constant (non-oscillatory). For our synchronization application we convert this to an oscillatory form with both a temporal and spatial periodicity. We also introduce a viscosity term to take account of dissipative processes that are important in most laboratory studies of dusty plasmas~\citep{Nakamura_PRL_1999, Nakamura_POP_2001}. The resulting equation has the form of a forced Korteweg-de Vries-Burgers (fKdV-B) equation. This equation provides a proper theoretical framework for the study of synchronization in a realistic dispersive plasma system that includes natural growth and dissipation of waves. 
\paragraph*{}
Numerical solution of this model equation, show clear signatures of harmonic (1:1)  and super-harmonic (1:2) synchronization. The characteristic features of the synchronization are delineated using power spectral density (PSD) plots, phase space plots  and Lissajous plots obtained from the time-series data collected at one spatial location. A parametric plot in the form of an Arnold tongue diagram shows multiple tongues, each corresponding to the existence region of a harmonic or a higher-order super-harmonic state. The tongues also show a branching behaviour. 
\paragraph*{}
The rest of the paper is organized as follows.  Section~\ref{fKdVB_model} briefly describes the fKdV-B model and the numerical approach adopted to solve it. The section also presents some numerical results for the undriven KdV and KdV-B equations as background information on the characteristic nonlinear features of the waves and to describe the diagnostic tools to be used for identifying synchronization phenomena. Section~\ref{sync_res} presents our main results on harmonic and super-harmonic synchronization using the fKdV-B model. A brief summary and some concluding discussion are provided in section~\ref{sum_con}.
%%%%%%%%%%%%%%%%%%%%%%%%%%%%%%%%%%%%%%%%%%%%%%%%%%%%%%%%%%%%%%%%%%%%%%%%%%%%%%%%%%%%%%%%%%%%%%%%%%%%%%%%%%%%%%%%%%%%%%%%%%%%%%%%%%%%%%%%%%%%%%%%%%%%%%%%%%%%%%%%%%%%%%%%%%%%%%%%%%%%%%%%%%%%%%%%%%%%%%%%%%%%%%
\section{The fKdV-B equation and the numerical approach}
\label{fKdVB_model}
%%%%%%%%%%%%%%%%%%%%%%%%%%%%%%%%%%%%%%%%%%%%%%%%%%%%%%%%%%%%%%%%%%%%%%%%%%%%%%%%%%%%%%%%%%%%%%%%%%%%%%%%%%%%%%%%%%%%%%%%%%%%%%%%%%%%%%%%%%%%%%%%%%%%%%%%%%%%%%%%%%%%%%%%%%%%%%%%%%%%%%%%%%%%%%%%%%%%%%%%%%%%%%
\paragraph*{}
The fKdV-B equation, a one-dimensional driven nonlinear partial differential equation, is of the form:
%%%%%%%%%%%%%%%%%%%%%%%%%%%%%%%%%%%%%
\begin{eqnarray}
&& 
\frac{\partial n (x,t)}{\partial t}
+\ \beta n(x,t)  \frac{\partial n(x,t)}{\partial x} 
\nonumber \\
&& 
+\ \gamma \frac{\partial^3 n(x,t)}{\partial x^3} 
-\ \eta \frac{\partial^2 n(x,t)}{\partial x^2} 
=\ F_s(x,t).
\label{Eqn_fKdVB_form}
\end{eqnarray}
%%%%%%%%%%%%%%%%%%%%%%%%%%%%%%%%%%%%%
%%%%%%%%%%%%%%%%%%%%%%%%%%%%%%%%%%%%%
\begin{figure} [h!]
\includegraphics[width = \columnwidth]{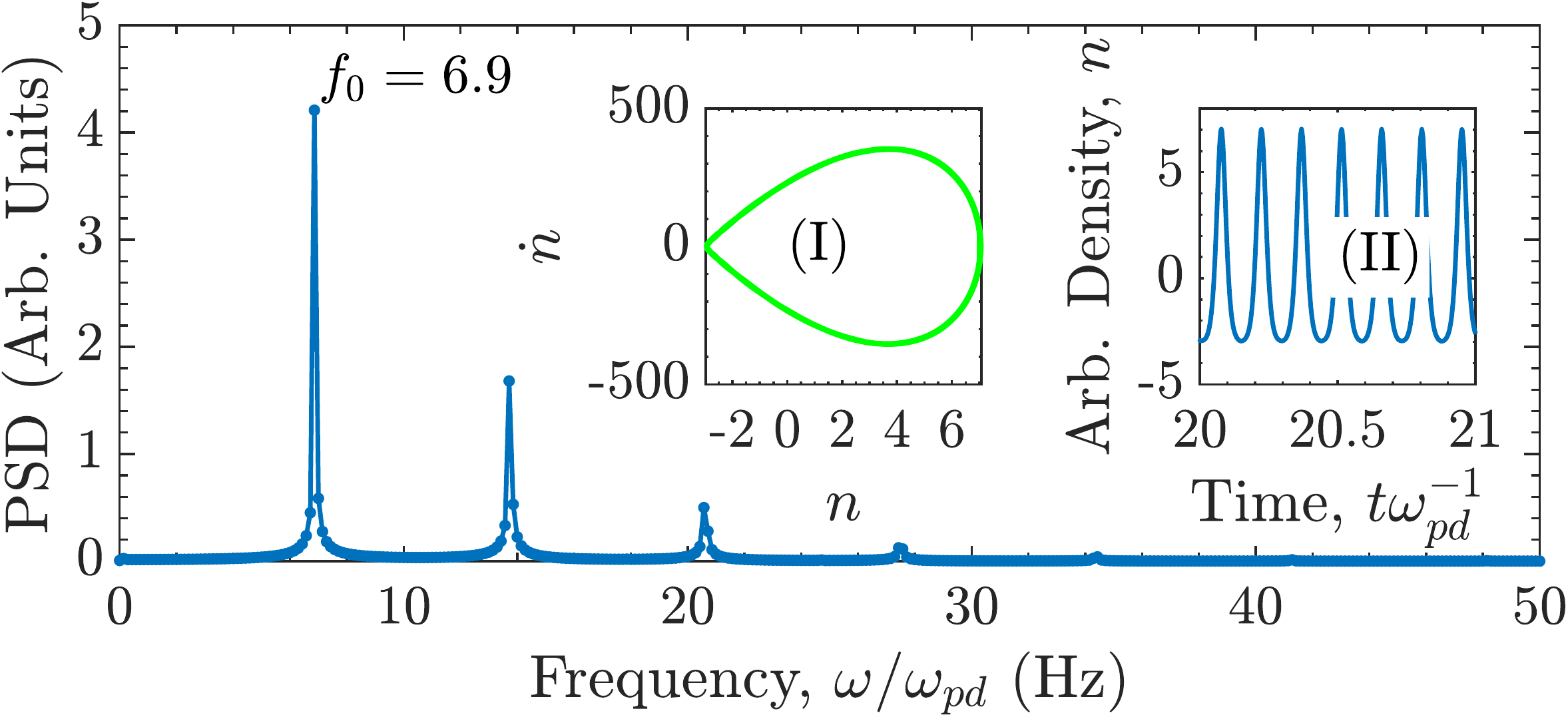}
\caption{
PSD for the times-series of KdV equation with initial perturbation Eq.~\eqref{Eqn_initial_form}.
Inset (II) shows the time-series obtained from the evolution of the KdV equation. Inset (I) is the phase space plot for the same time-series.}
\label{Fig_1}
\end{figure}
%%%%%%%%%%%%%%%%%%%%%%%%%%%%%%%%%%%%% 
%%%%%%%%%%%%%%%%%%%%%%%%%%%%%%%%%%%%%
\begin{figure} [h!]
\includegraphics[width = \columnwidth]{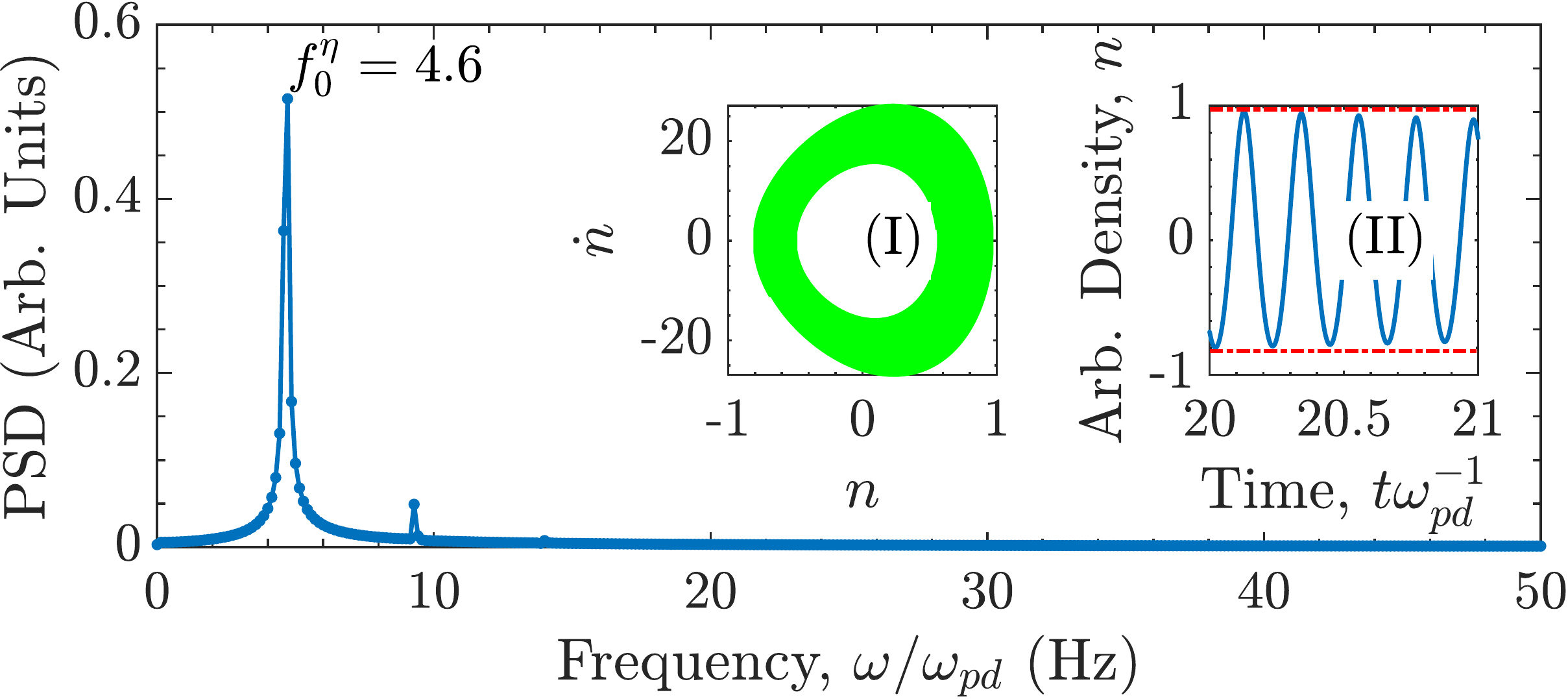}
\caption{ 
PSD of the evolution of KdV-B equation with initial perturbation Eq.~\eqref{Eqn_initial_form} at $\eta=0.0025$.
The time-series of the KdV-B equation is shown in the inset (II). Inset (I) is the phase space plot of the same KdV-B time-series.}
\label{Fig_2}
\end{figure}
%%%%%%%%%%%%%%%%%%%%%%%%%%%%%%%%%%%%%
%%%%%%%%%%%%%%%%%%%%%%%%%%%%%%%%%%%%%
\begin{figure*} [ht!]
\includegraphics[width = \textwidth]{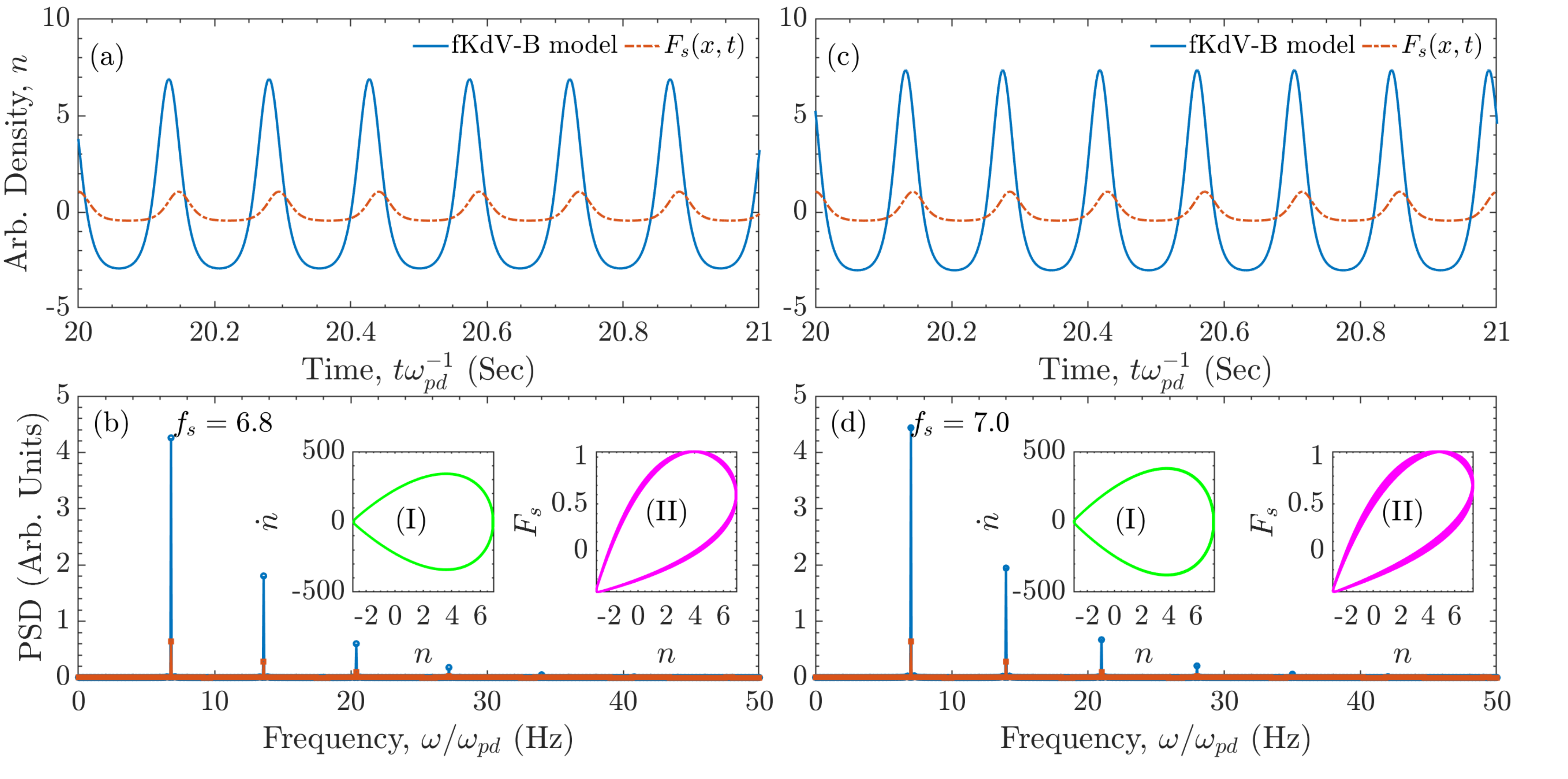}
\caption{ 
The harmonic (1:1) synchronization in the fKdV-B model with $f_s < f_0$ and $f_s > f_0$.
The time-series of the fKdV-B model (blue bold line) and the forcing (brown dash-dotted line) at driver frequency (a) $f_s = 6.8$ Hz with threshold amplitude $A_s = 0.15A_0$  and (b) $f_s = 7.0$ Hz with threshold amplitude $A_s = 0.15A_0$. (c) PSD of times-series (a). (d) PSD of time-series (b).
The insets (I) phase space plot and (II) Lissajous figure reflect the frequency locking at the driver frequency.
}
\label{Fig_3}
\end{figure*}
%%%%%%%%%%%%%%%%%%%%%%%%%%%%%%%%%%%%%
Here $n(x,t)$ is the dependent variable (the perturbed density in this case) and $F_s(x,t)$ is an external spatio-temporal forcing term. $\beta$, $\gamma$, and $\eta$ are positive quantities representing the strength of nonlinearity, dispersion, and viscous damping, respectively. The spatial coordinate $x$ and time $t$ are normalized by the plasma Debye length $\lambda_D$ and the dust plasma period $\omega_{pd}^{-1}$, respectively. 
\paragraph*{}
The fKdV-B equation can be derived from the full fluid-Poisson set of equations, in the weakly nonlinear, dispersive and dissipative regime by using a reductive perturbation method~\citep{Nakamura_PRL_1999, Veeresha_PRE_2010}. Such a derivation in the absence of the viscosity term has been given in detail by Sen~{\it et al.}~\cite{Sen_ASR_2015}. In their derivation the source term was taken to be a constant; here it is replaced by a more general term that is a periodic function in both space and time. The KdV-B equation (\textit{i.e.,} Eq.~\eqref{Eqn_fKdVB_form} in the absence of the driving term) is well known in the literature~\citep{Nakamura_PRL_1999, Veeresha_PRE_2010, Nakamura_POP_2001} and has been employed in the past to model oscillatory shocks in dusty plasmas~\cite{Nakamura_PRL_1999,Jaiswal_POP_2016}. The model has also been used to study temporal chaos or spatial chaos by using a randomly time varying~\citep{Orowski_PRE_1994} or randomly space varying~\citep{Rech_EPJB_2013} driving term. In this work, we use a spatio-temporally varying periodic source and carry out a numerical investigation of Eq.~\eqref{Eqn_fKdVB_form} to study the synchronization of DAWs. 
\paragraph*{}
The driving source is taken to be of the form of a cnoidal-square travelling wave, 
%%%%%%%%%%%%%%%%%%%%%%%%%%%%%%%%%%%%%
\begin{equation}
  F_s(x,t) = A_s cn^2[2 K(\kappa_s) \{ x/\lambda_s - f_s t \};\ \kappa_s] 
  \label{Eqn_forcing_form}
\end{equation}
%%%%%%%%%%%%%%%%%%%%%%%%%%%%%%%%%%%%%
where $cn$ is the Jacobi elliptic function, $A_s$ is the driving amplitude, \textcolor{black}{$\lambda_s$ is the spatial wave length} and $f_s$ is the driving frequency. $K(\kappa)$ is the complete elliptical integral of first kind and the elliptic parameter $\kappa$ is a measure of the nonlinearity of the wave. 
The cnoidal-square travelling wave is an exact solution of the KdV equation (\textit{i.e.,} a solution of the model equation in the absence of viscosity and the driving term). It can therefore mimic the driving of the system by a DAW arising from an external (coupled) plasma source. For the numerical solution of Eq.~\eqref{Eqn_fKdVB_form} the initial waveform is also taken to be of the form,
%%%%%%%%%%%%%%%%%%%%%%%%%%%%%%%%%%%%%
\begin{equation}
 n(x,t=0) = A_0 cn^2[2 K(\kappa_0) \{ x/\lambda_0\};\ \kappa_0].
 \label{Eqn_initial_form}
\end{equation}
%%%%%%%%%%%%%%%%%%%%%%%%%%%%%%%%%%%%%
with the values of $A_0$, $f_0$ and $\lambda_0$ different from those of the driving source. The idea is to see whether the final driven modes of the system synchronize to the frequency of the driver. Equation (\ref{Eqn_fKdVB_form}) is solved for various values of $f_s$ and $A_s$ in order to find the regions of synchronization in the parameter space of $(A_s,f_s)$. 
%%%%%%%%%%%%%%%%%%%%%%%%%%%%%%%%%%%%%
\begin{figure*} [ht!]
\includegraphics[width = \textwidth]{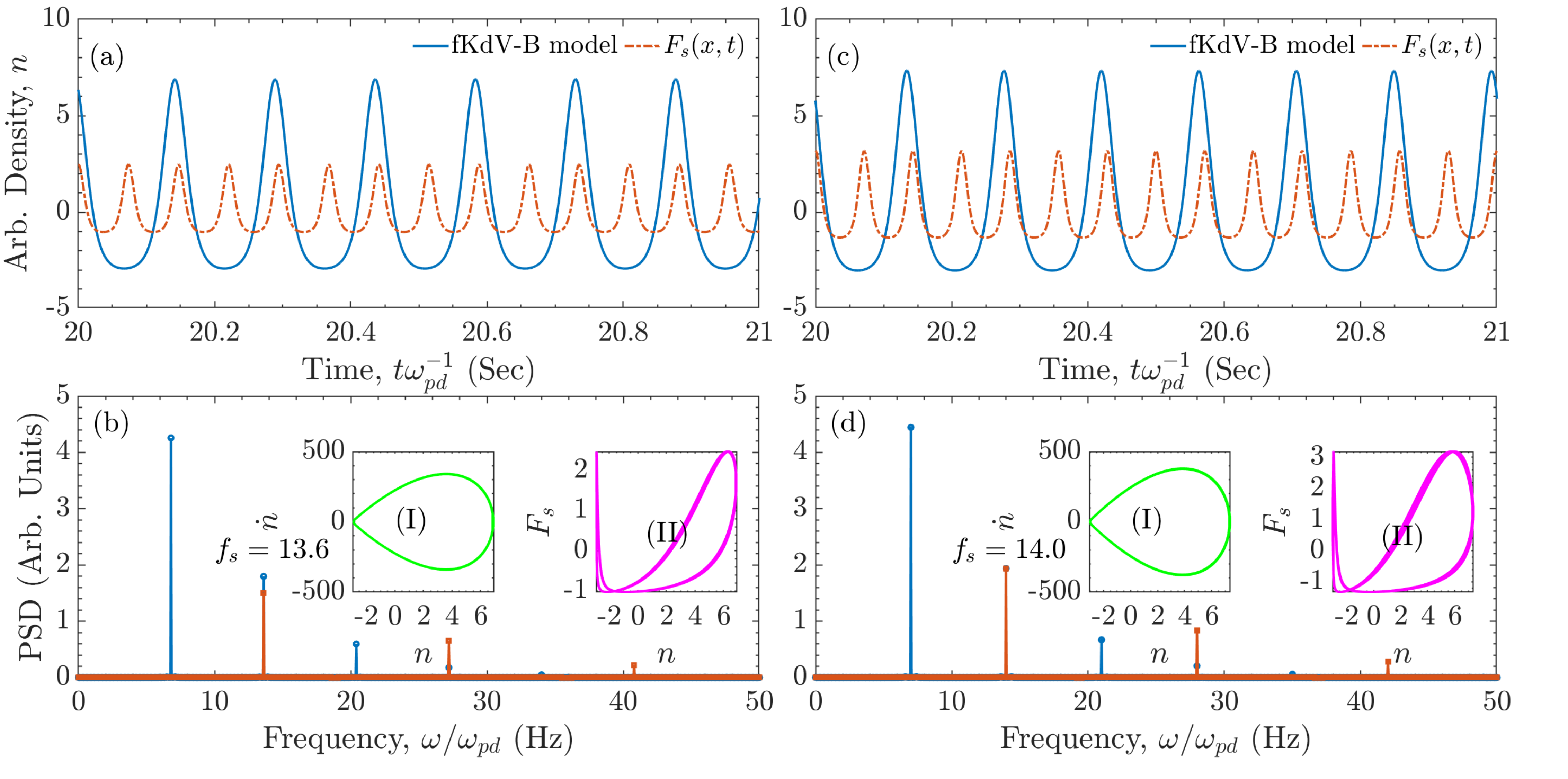}
\caption{ 
The super-harmonic (1:2) synchronization in the fKdV-B model with $f_s < 2f_0$ and $f_s > 2f_0$.
The time-series of the fKdV-B model (blue bold line) and the forcing (brown dash-dotted line) at driver frequency (a) $f_s = 13.6$ Hz with threshold amplitude $A_s = 0.35A_0$ and (b) $f_s = 14.0$ Hz with threshold amplitude $A_s = 0.45A_0$. (c) PSD of times-series (a). (d) PSD of time-series (b).
The insets (I) phase space plot and (II) Lissajous figure reflect the frequency locking at half of the driver frequency.
}
\label{Fig_4}
\end{figure*}
%%%%%%%%%%%%%%%%%%%%%%%%%%%%%%%%%%%%%
\paragraph*{}
Our numerical investigation of the fKdV-B equation is based on the pseudo-spectral method~\citep{Boyd_DP_2003} and uses periodic boundary conditions. The code is first bench marked by reproducing earlier results~\cite{Sen_ASR_2015, Ajaz_POP_2022} obtained for the fKdV equation. The various parameter values associated with the model are taken to be as follows: The Jacobi elliptic parameters $\kappa_0 = \kappa_s = 0.98$ for Eqs.~\eqref{Eqn_forcing_form} and~\eqref{Eqn_initial_form}. The wave vector of the initial perturbation \textit{i.e.,} $k_0 = 16 k_m$ where $k_m = (2\pi)/L_x$ being the minimum wave vector associated with a system of length $L_x = 6\pi$. The corresponding wavelength \textit{i.e.,} $\lambda_0 = (2\pi)/ k_0$ and amplitude $A_0 = 10$ are kept fixed throughout the analysis. We have taken $k_s = 16 k_m$ and $k_s = 2 \times 16 k_m$ for studying harmonic (1:1), and super-harmonic (1:2) synchronization states. The corresponding forcing wavelength is $\lambda_s = (2\pi)/ k_s$. Throughout the analysis, we have only varied the forcing amplitude, $A_s$ and forcing frequency, $f_s$. We have also chosen the coefficient values of the model equation as  $\beta = 2.5$, $\gamma = 0.0667$ and $\eta = 0.0025$ for all the runs. Based on the chosen parameters $\alpha$, $\beta$, $\kappa_0$ and $k_0$, the value of the initial amplitude $(A_0=10)$ is derived using the relationship provided in Mir \textit{et al.}~\cite{Ajaz_POP_2020}. The amplitude of the initial perturbation chosen in this fashion will be governed by the exact solution of the KdV and will be a stable solution of KdV for this particular amplitude.
\paragraph*{}
We evolve the initial perturbation in  Eq.~\eqref{Eqn_fKdVB_form} over long times for these various different parameter values.  During the spatio-temporal evolution, we collect a time-series of the density field at a fixed spatial location and use it to calculate the power spectral density. The PSD provides a useful tool for distinguishing between synchronized and un-synchronized states. 
\paragraph*{}
As an illustrative example, we show in Fig.~\ref{Fig_1} the PSD, the time-series and the phase space plot of the solution, obtained for a pure KdV equation (\textit{i.e.,} for $\eta=A_s=0$). The time-series data has been collected up to  $t_{max} = 40$ $\omega_{pd}^{-1}$ with a time step $dt = 10^{-5}$ $\omega_{pd}^{-1}$. The maximum sampling frequency $f_{S} = 1/dt$ and the Nyquist frequency is $f_N = f_{S}/2$. This leads to a frequency resolution of $df = 1/t_{max}$ for the collected time-series. The time-series data corresponding to the first 20 periods is discarded to remove transient effects while constructing the PSD.  
In Fig.~\ref{Fig_1} the nonlinear character of the mode is evident from the presence of the higher harmonics in the PSD and from the shape of wave form in the time-series. The single cycle phase space plot with its form resembling a separatrix curve indicates an undamped nonlinear periodic wave, in this case the exact cnoidal-square wave. 
\paragraph*{}
For comparison, we present in Fig.~\ref{Fig_2} the corresponding results for the undriven KdV-B equation (with $\eta =0.0025$ and $F_s=0$). The effect of viscous damping is seen in the frequency shift of the fundamental component in the PSD towards a lower value of $f_0^{\eta} = 4.6$ Hz, the reduced amplitude in the time-series and the spiralling of the phase space plot towards the origin. It is clear that in the presence of finite viscosity the cnoidal-square wave can no longer be sustained as a nonlinear solution of Eq.~\eqref{Eqn_fKdVB_form} with $F_s = 0$ and the initial perturbation decays in time. The question is whether by driving the system with a periodic source one can revive and sustain a nonlinear solution that is also synchronized with the driver.  The answer is in the positive and we next present our results on such a phenomenon.
%%%%%%%%%%%%%%%%%%%%%%%%%%%%%%%%%%%%%
\begin{figure*} [ht!]
\includegraphics[width = \columnwidth, height= \columnwidth]{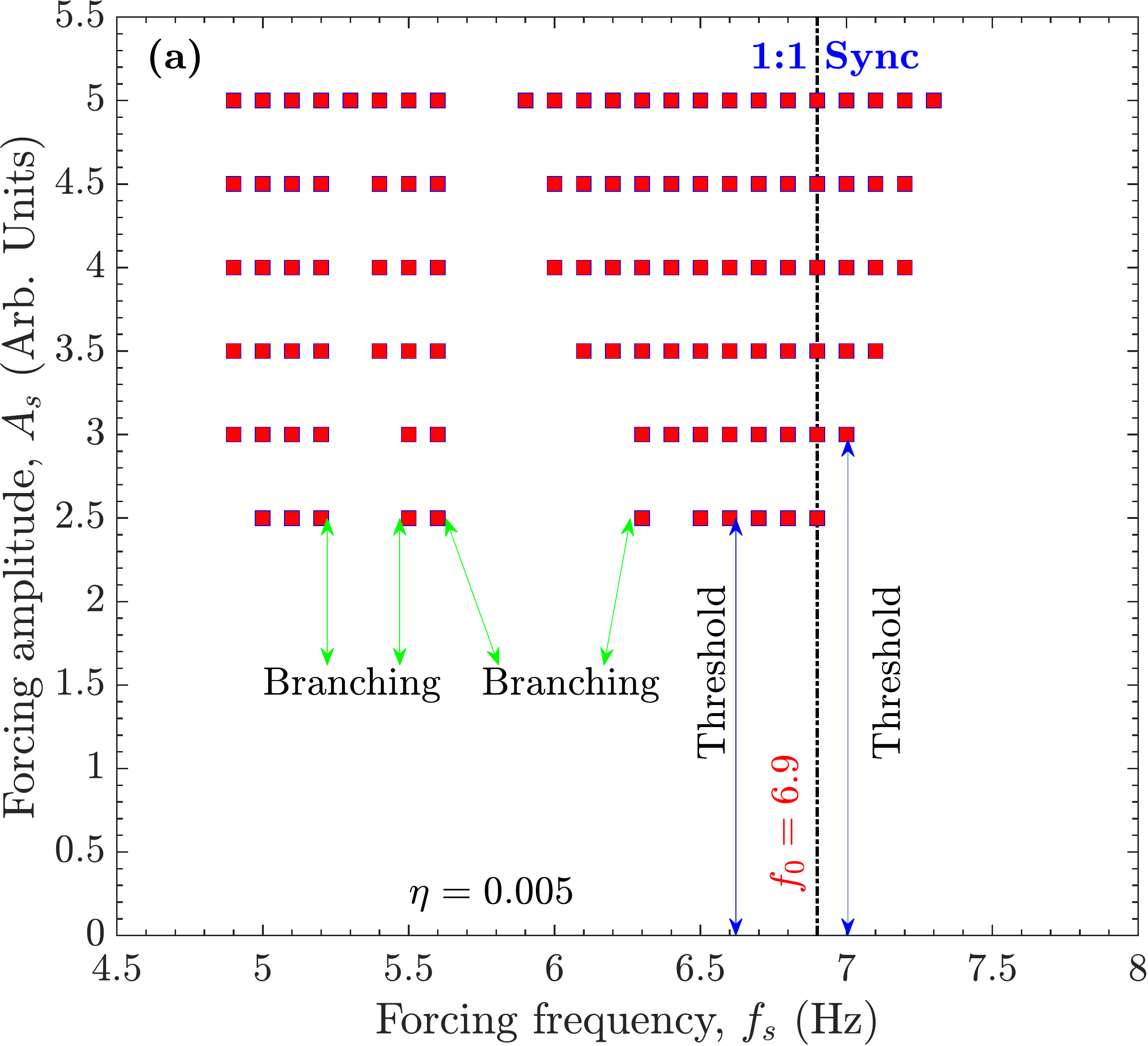}
\includegraphics[width = \columnwidth, height= \columnwidth]{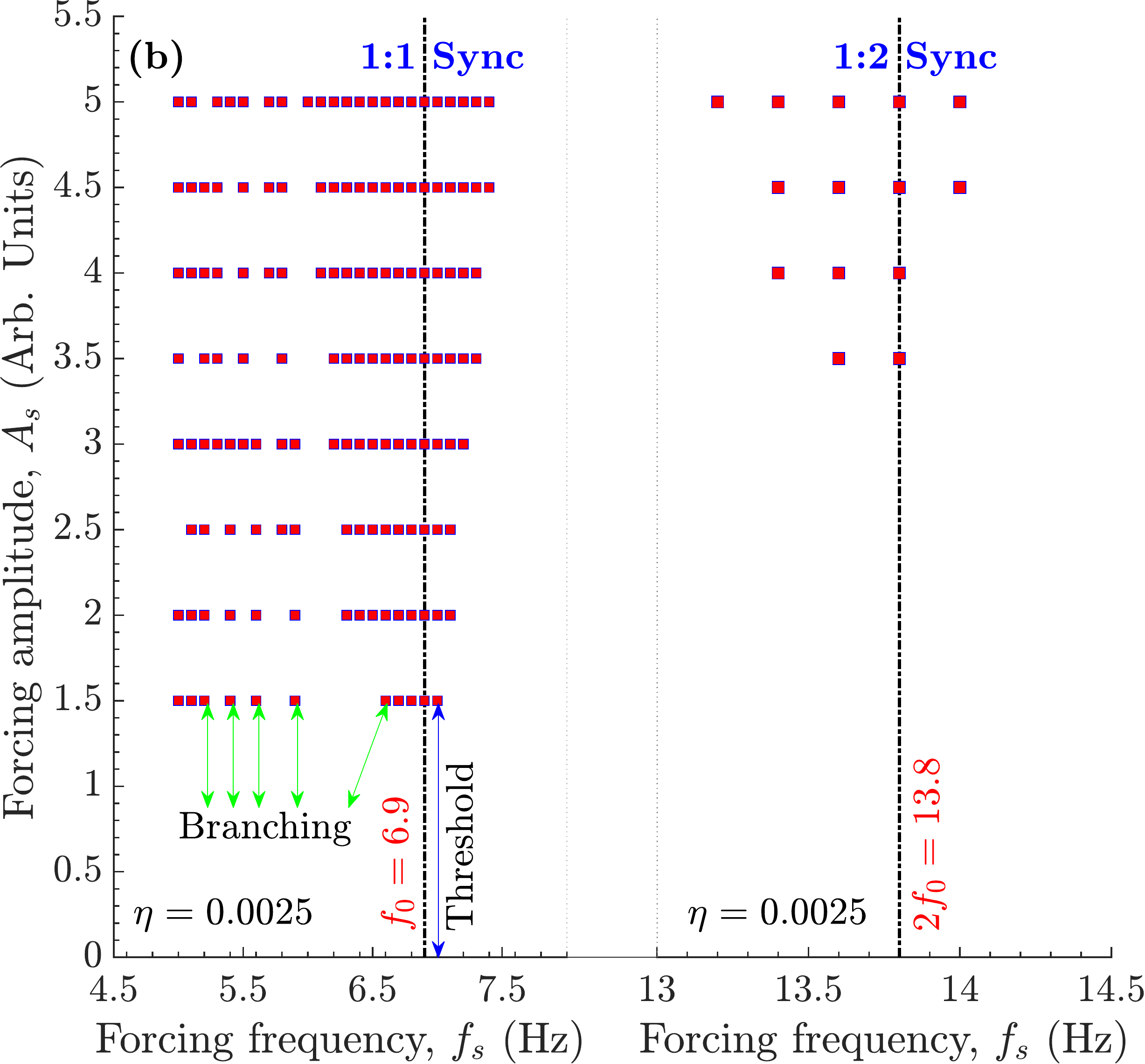}
\caption{ 
The Arnold tongue diagrams for (a) harmonic (1:1) synchronization state with $\eta=0.005$ and (b) harmonic (1:1) and super-harmonic (1:2) synchronization states with $\eta = 0.0025$ in the fKdV-B model. 
In subplot (a) the amplitude is varied from $A_s = 0.25A_0$ to $A_s = 0.50A_0$ for 1:1 synchronization. In subplot (b) the amplitude is varied from $A_s = 0.15A_0$ to $A_s = 0.50A_0$ for 1:1, and $A_s = 0.35A_0$ to $A_s = 0.50A_0$ for 1:2 synchronization.
}
\label{Fig_5}
\end{figure*}
%%%%%%%%%%%%%%%%%%%%%%%%%%%%%%%%%%%%%
%%%%%%%%%%%%%%%%%%%%%%%%%%%%%%%%%%%%%%%%%%%%%%%%%%%%%%%%%%%%%%%%%%%%%%%%%%%%%%%%%%%%%%%%%%%%%%%%%%%%%%%%%%%%%%%%%%%%%%%%%%%%%%%%%%%%%%%%%%%%%%%%%%%%%%%%%%%%%%%%%%%%%%%%%%%%%%%%%%%%%%%%%%%%%%%%%%%%%%%%%%%%%%
\section{Synchronization in fKdV-B model}
\label{sync_res}
%%%%%%%%%%%%%%%%%%%%%%%%%%%%%%%%%%%%%%%%%%%%%%%%%%%%%%%%%%%%%%%%%%%%%%%%%%%%%%%%%%%%%%%%%%%%%%%%%%%%%%%%%%%%%%%%%%%%%%%%%%%%%%%%%%%%%%%%%%%%%%%%%%%%%%%%%%%%%%%%%%%%%%%%%%%%%%%%%%%%%%%%%%%%%%%%%%%%%%%%%%%%%%
In this section we present the main results of our work, namely, the synchronization of the solutions of Eq.~\eqref{Eqn_fKdVB_form} to an external driver of the form given by Eq.~\eqref{Eqn_forcing_form}. We begin by discussing harmonic (1:1) synchronization  for which we choose the driving frequency to be slightly away from the fundamental frequency of $f_0=6.9$ Hz that is characteristic fundamental frequency of the undriven system. Two cases are considered, namely, $f_s = 6.8$ and $f_s = 7.0$ Hz. The driving amplitude in both cases is taken to be $A_s=15$ with $A_0 = 10$ and $\eta=0.0025$. Figure~\ref{Fig_3} shows the attainment of harmonic (1:1) synchronization for both these cases with subplots (a, b) devoted to $f_s= 6.8$ Hz and (c, d) to $f_s= 7.0$ Hz, respectively. As can be seen from the time-series plots in (a) and (c) the driven solutions are indeed locked to the driver. This is also clearly seen in the PSDs where the fundamental frequencies of the driven solutions are indeed at the frequency of the driver. Furthermore, the phase space plots in (b) and (d) show that these solutions constitute undamped nonlinear periodic waves that are maintained by a balance between the nonlinear steepening, dispersive broadening, viscous damping and amplification due to the external pumping by the driving term. The resultant phase space curve, that has the characteristic shape of a separatrix, represents a stationary  cnoidal wave solution. The presence of dissipation seems to be necessary for sustaining this synchronized driven solution.
\textcolor{black}{We have found that in the partial differential equation  Eq.~(\ref{Eqn_fKdVB_form}), including not just nonlinear and dispersive terms, but also a linear dissipative term, allowed achieving synchronization of a wave. When we turned off dissipation, by setting the viscosity coefficient to zero in Eq.~(\ref{Eqn_fKdVB_form}), we did not observe synchronization of the wave, for the conditions that we studied here. This is different from the case of a point oscillator, as described by the van der Pol oscillator Eq.~(\ref{fVdP_eqn}), which requires a nonlinear dissipation term to obtain synchronization.}  In the absence of viscosity one only gets nonlinear mixing from the model as has been reported earlier in Mir {\it et al.}~\cite{Ajaz_POP_2020, Ajaz_POP_2022}. \textcolor{black}{The amount of viscosity also determines the threshold condition for the driver amplitude.}
\paragraph*{}
To explore super-harmonic (1:2) synchronization we again consider two cases of $f_s= 13.6$ Hz and $f_s= 14$ Hz which are slightly below and above the first harmonic frequency $2f_0 = 13.8$ Hz of the undriven system.
The results are shown in Fig.~\ref{Fig_4} where the subplots (a,c) are devoted to $f_s=13.6$ Hz and (b,d) to $f_s=14$ Hz respectively. As in the previous case of harmonic synchronization, we see clear evidence of super-harmonic (2:1) synchronization in the time-series plots, the PSDs and the phase space plots. The Lissajous figures have a number eight-like trajectory which is indicative of a (1:2) synchronized state. One significant difference from the harmonic synchronization case is that the minimum threshold amplitude for the driver to achieve a 1:2 state is different for the cases $f_s< 2f_0$ and $f_s>2f_0$. They are $A_s = 0.35 A_0$ and $A_s = 0.45 A_0$ respectively.
\paragraph*{}
Finally, in Fig.~\ref{Fig_5} we present a consolidated picture of the existence domain of these synchronized states in the parameter space of the driver frequency $f_s$  and driver amplitude $A_s$ in form of an Arnold tongue diagram.
To obtain the Arnold tongue diagram, $A_s$ is varied in steps of 0.5 (which is $0.05 A_0$) from 0 to 5 while $f_s$ is varied in steps of 0.1 Hz for harmonic synchronization and 0.2 Hz for the super-harmonic case. Figure~\ref{Fig_5}(a) shows the 1:1 entrained state tongue for $\eta = 0.005$ and Fig.~\ref{Fig_5}(b) shows the 1:1 and 1:2 entrained state tongues for $\eta = 0.0025$. 
\paragraph*{}
We observe several interesting features in the Arnold tongue diagrams. To start with, there is always a threshold amplitude $A_s$ below which no synchronization occurs. 
For the harmonic (1:1) synchronization it is $A_s = 0.25A_0 = 2.5$ for $\eta = 0.005$, while it is $A_s = 0.15 A_0 = 1.5$ for $\eta =0.0025$.
\textcolor{black}{This is unlike the harmonic synchronization phenomenon observed in a driven van der Pol model where no such threshold is found~\cite{Jensen_AJP_2002}.} Another important feature is a distinctive branching of the Arnold tongue that is clearly seen for the (1:1) states at low forcing amplitudes. Arrows mark this branching both in Fig.~\ref{Fig_5}(a) and Fig.~\ref{Fig_5}(b). The branching gives rise to a non-synchronized region between the frequencies $f_s= 6.4$ Hz to $f_s=5.7$ Hz at driver amplitude $A_s=0.25A_0 = 2.5$ in Fig.~\ref{Fig_5}(a). This branching narrows down with the increase in $A_s$. Another branch is seen in Fig.~\ref{Fig_5}(a) between $f_s=5.4$ Hz and $f_s=5.3$ Hz which also narrows down with  increase in $A_s$. A third feature is the asymmetric nature of the tongue structures about $f_0$. The frequency width over which synchronization can be obtained is much broader for $f_s < f_0$ compared to $f_s > f_0$.
%%%%%%%%%%%%%%%%%%%%%%%%%%%%%%%%%%%%%%%%%%%%%%%%%%%%%%%%%%%%%%%%%%%%%%%%%%%%%%%%%%%%%%%%%%%%%%%%%%%%%%%%%%%%%%%%%%%%%%%%%%%%%%%%%%%%%%%%%%%%%%%%%%%%%%%%%%%%%%%%%%%%%%%%%%%%%%%%%%%%%%%%%%%%%%%%%%%%%%%%%%%%%%
\section{Summary and Conclusions}
\label{sum_con}
%%%%%%%%%%%%%%%%%%%%%%%%%%%%%%%%%%%%%%%%%%%%%%%%%%%%%%%%%%%%%%%%%%%%%%%%%%%%%%%%%%%%%%%%%%%%%%%%%%%%%%%%%%%%%%%%%%%%%%%%%%%%%%%%%%%%%%%%%%%%%%%%%%%%%%%%%%%%%%%%%%%%%%%%%%%%%%%%%%%%%%%%%%%%%%%%%%%%%%%%%%%%%%
\paragraph*{}
To summarize, we have studied the phenomenon of synchronization of dust acoustic waves to an external periodic driver in a model system described by the forced Korteweg-de Vries-Burgers equation. 
This equation provides a proper theoretical framework and a better physical model compared to the van der Pol oscillator model for studying  the dynamics of nonlinear dust acoustic waves by properly accounting for nonlinear, dispersive and dissipative influences on the waves.  Using the model, we have successfully demonstrated harmonic (1:1) and super-harmonic (1:2) synchronization states of DAWs. Our numerical findings show a strong qualitative resemblance to the experimental results on synchronization reported by Ruhunusiri~\textit{et al.}~\citep{Suranga_PRE_2012}. In particular, comparison of our theoretical Arnold tongue diagram with their experimental one shows the following common features. As in the experimental Arnold tongue diagram we see the existence of amplitude thresholds as well as clear evidence of the branching phenomena. However there are also important differences. With our model we have not been able to obtain sub-harmonic synchronization that have been observed in the experiment. Furthermore, our model uses an external driver that closely resembles a nonlinear natural mode of the system whereas in the experiment a purely time varying external sinusoidal driver has been used. However, it is not clear what form this driver takes inside the plasma system and whether it manifests itself as a spatio-temporally varying perturbation. These and other questions, such as the absence of sub-harmonic synchronization in the equation, \textcolor{black}{the neglect of dissipation arising from gas friction on the dust particles, \textit{etc.},} remain to be explored in order to further improve the model.
%%%%%%%%%%%%%%%%%%%%%%%%%%%%%%%%%%%%%%%%%%%%%%%%%%%%%%%%%%%%%%%%%%%%%%%%%%%%%%%%%%%%%%%%%%%%%%%%%%%%%%%%%%%%%%%%%%%%%%%%%%%%%%%%%%%%%%%%%%%%%%%%%%%%%%%%%%%%%%%%%%%%%%%%%%%%%%%%%%%%%%%%%%%%%%%%%%%%%%%%%%%%%%
\begin{acknowledgements}
Work done by ST and AM was supported by the Indian Institute of Technology Jammu Seed Grant No. SG0012.
AS is thankful to the Indian National Science Academy (INSA) for the INSA Honorary Scientist position.
CC and GG acknowledge NASA-JPL subcontract No. 1573108 and NRL Base Funds.
JG was supported by United States Department of Energy Grant No. DE-SC0014566, NASA/JPL RSA No. 1672641, and National Science Foundation Grant No. PHY-1740379.
\end{acknowledgements}
%%%%%%%%%%%%%%%%%%%%%%%%%%%%%%%%%%%%%%%%%%%%%%%%%%%%%%%%%%%%%%%%%%%%%%%%%%%%%%%%%%%%%%%%%%%%%%%%%%%%%%%%%%%%%%%%%%%%%%%%%%%%%%%%%%%%%%%%%%%%%%%%%%%%%%%%%%%%%%%%%%%%%%%%%%%%%%%%%%%%%%%%%%%%%%%%%%%%%%%%%%%%%%
\bibliography{Sync_Plasma}

\begin{thebibliography}{51}%
\makeatletter
\providecommand \@ifxundefined [1]{%
 \@ifx{#1\undefined}
}%
\providecommand \@ifnum [1]{%
 \ifnum #1\expandafter \@firstoftwo
 \else \expandafter \@secondoftwo
 \fi
}%
\providecommand \@ifx [1]{%
 \ifx #1\expandafter \@firstoftwo
 \else \expandafter \@secondoftwo
 \fi
}%
\providecommand \natexlab [1]{#1}%
\providecommand \enquote  [1]{``#1''}%
\providecommand \bibnamefont  [1]{#1}%
\providecommand \bibfnamefont [1]{#1}%
\providecommand \citenamefont [1]{#1}%
\providecommand \href@noop [0]{\@secondoftwo}%
\providecommand \href [0]{\begingroup \@sanitize@url \@href}%
\providecommand \@href[1]{\@@startlink{#1}\@@href}%
\providecommand \@@href[1]{\endgroup#1\@@endlink}%
\providecommand \@sanitize@url [0]{\catcode `\\12\catcode `\$12\catcode
  `\&12\catcode `\#12\catcode `\^12\catcode `\_12\catcode `\%12\relax}%
\providecommand \@@startlink[1]{}%
\providecommand \@@endlink[0]{}%
\providecommand \url  [0]{\begingroup\@sanitize@url \@url }%
\providecommand \@url [1]{\endgroup\@href {#1}{\urlprefix }}%
\providecommand \urlprefix  [0]{URL }%
\providecommand \Eprint [0]{\href }%
\providecommand \doibase [0]{http://dx.doi.org/}%
\providecommand \selectlanguage [0]{\@gobble}%
\providecommand \bibinfo  [0]{\@secondoftwo}%
\providecommand \bibfield  [0]{\@secondoftwo}%
\providecommand \translation [1]{[#1]}%
\providecommand \BibitemOpen [0]{}%
\providecommand \bibitemStop [0]{}%
\providecommand \bibitemNoStop [0]{.\EOS\space}%
\providecommand \EOS [0]{\spacefactor3000\relax}%
\providecommand \BibitemShut  [1]{\csname bibitem#1\endcsname}%
\let\auto@bib@innerbib\@empty
%</preamble>
\bibitem [{\citenamefont {Pikovsky}\ \emph {et~al.}(2001)\citenamefont
  {Pikovsky}, \citenamefont {Rosenblum},\ and\ \citenamefont
  {Kurths}}]{Pikovsky_CUP_2001}%
  \BibitemOpen
  \bibfield  {author} {\bibinfo {author} {\bibfnamefont {A.}~\bibnamefont
  {Pikovsky}}, \bibinfo {author} {\bibfnamefont {M.}~\bibnamefont {Rosenblum}},
  \ and\ \bibinfo {author} {\bibfnamefont {J.}~\bibnamefont {Kurths}},\ }\href
  {\doibase 10.1017/CBO9780511755743} {\emph {\bibinfo {title}
  {Synchronization: A Universal Concept in Nonlinear Sciences}}}\ (\bibinfo
  {publisher} {Cambridge University Press, {Cambridge}},\ \bibinfo {year}
  {2001})\BibitemShut {NoStop}%
\bibitem [{\citenamefont {Strogatz}(2004)}]{Strogatz_PP_2004}%
  \BibitemOpen
  \bibfield  {author} {\bibinfo {author} {\bibfnamefont {S.}~\bibnamefont
  {Strogatz}},\ }\href@noop {} {\emph {\bibinfo {title} {Sync: The Emerging
  Science of Spontaneous Order}}}\ (\bibinfo  {publisher} {Penguin, UK},\
  \bibinfo {year} {2004})\BibitemShut {NoStop}%
\bibitem [{\citenamefont {Balanov}\ \emph {et~al.}(2009)\citenamefont
  {Balanov}, \citenamefont {Janson}, \citenamefont {Postnov},\ and\
  \citenamefont {Sosnovtseva}}]{Balanov_SV_2009}%
  \BibitemOpen
  \bibfield  {author} {\bibinfo {author} {\bibfnamefont {A.}~\bibnamefont
  {Balanov}}, \bibinfo {author} {\bibfnamefont {N.}~\bibnamefont {Janson}},
  \bibinfo {author} {\bibfnamefont {D.}~\bibnamefont {Postnov}}, \ and\
  \bibinfo {author} {\bibfnamefont {O.}~\bibnamefont {Sosnovtseva}},\
  }\href@noop {} {\emph {\bibinfo {title} {Synchronization: From Simple to
  Complex}}}\ (\bibinfo  {publisher} {Springer-Verlag, {Berlin}},\ \bibinfo
  {year} {2009})\BibitemShut {NoStop}%
\bibitem [{\citenamefont {Acebr\'on}\ \emph {et~al.}(2005)\citenamefont
  {Acebr\'on}, \citenamefont {Bonilla}, \citenamefont {P\'erez~Vicente},
  \citenamefont {Ritort},\ and\ \citenamefont {Spigler}}]{Acebron_RMP_2005}%
  \BibitemOpen
  \bibfield  {author} {\bibinfo {author} {\bibfnamefont {J.~A.}\ \bibnamefont
  {Acebr\'on}}, \bibinfo {author} {\bibfnamefont {L.~L.}\ \bibnamefont
  {Bonilla}}, \bibinfo {author} {\bibfnamefont {C.~J.}\ \bibnamefont
  {P\'erez~Vicente}}, \bibinfo {author} {\bibfnamefont {F.}~\bibnamefont
  {Ritort}}, \ and\ \bibinfo {author} {\bibfnamefont {R.}~\bibnamefont
  {Spigler}},\ }\href {\doibase 10.1103/RevModPhys.77.137} {\bibfield
  {journal} {\bibinfo  {journal} {Rev. Mod. Phys.}\ }\textbf {\bibinfo {volume}
  {77}},\ \bibinfo {pages} {137} (\bibinfo {year} {2005})}\BibitemShut
  {NoStop}%
\bibitem [{\citenamefont {Kuramoto}(1984)}]{Kuramoto_PTPS_1984}%
  \BibitemOpen
  \bibfield  {author} {\bibinfo {author} {\bibfnamefont {Y.}~\bibnamefont
  {Kuramoto}},\ }\href@noop {} {\bibfield  {journal} {\bibinfo  {journal}
  {Progr. Theoret. Phys. Suppl.}\ }\textbf {\bibinfo {volume} {79}},\ \bibinfo
  {pages} {223} (\bibinfo {year} {1984})}\BibitemShut {NoStop}%
\bibitem [{\citenamefont {Aronson}\ \emph {et~al.}(1990)\citenamefont
  {Aronson}, \citenamefont {Ermentrout},\ and\ \citenamefont
  {Kopell}}]{Aronson_PDNP_1990}%
  \BibitemOpen
  \bibfield  {author} {\bibinfo {author} {\bibfnamefont {D.}~\bibnamefont
  {Aronson}}, \bibinfo {author} {\bibfnamefont {G.}~\bibnamefont {Ermentrout}},
  \ and\ \bibinfo {author} {\bibfnamefont {N.}~\bibnamefont {Kopell}},\ }\href
  {\doibase https://doi.org/10.1016/0167-2789(90)90007-C} {\bibfield  {journal}
  {\bibinfo  {journal} {Physica D: Nonlinear Phenomena}\ }\textbf {\bibinfo
  {volume} {41}},\ \bibinfo {pages} {403} (\bibinfo {year} {1990})}\BibitemShut
  {NoStop}%
\bibitem [{\citenamefont {Kuramoto}(2003)}]{Kuramoto_DP_2003}%
  \BibitemOpen
  \bibfield  {author} {\bibinfo {author} {\bibfnamefont {Y.}~\bibnamefont
  {Kuramoto}},\ }\href@noop {} {\emph {\bibinfo {title} {Chemical Oscillations,
  Waves, and Turbulence}}}\ (\bibinfo  {publisher} {Dover, New York},\ \bibinfo
  {year} {2003})\BibitemShut {NoStop}%
\bibitem [{\citenamefont {Williams}\ and\ \citenamefont
  {Duff}(2010)}]{Williams_POP_2010}%
  \BibitemOpen
  \bibfield  {author} {\bibinfo {author} {\bibfnamefont {J.~D.}\ \bibnamefont
  {Williams}}\ and\ \bibinfo {author} {\bibfnamefont {J.}~\bibnamefont
  {Duff}},\ }\href {\doibase 10.1063/1.3357335} {\bibfield  {journal} {\bibinfo
   {journal} {Phys. Plasmas}\ }\textbf {\bibinfo {volume} {17}},\ \bibinfo
  {pages} {033702} (\bibinfo {year} {2010})}\BibitemShut {NoStop}%
\bibitem [{\citenamefont {Block}\ \emph {et~al.}(2001)\citenamefont {Block},
  \citenamefont {Piel}, \citenamefont {Schr\"oder},\ and\ \citenamefont
  {Klinger}}]{Block_PRE_2001}%
  \BibitemOpen
  \bibfield  {author} {\bibinfo {author} {\bibfnamefont {D.}~\bibnamefont
  {Block}}, \bibinfo {author} {\bibfnamefont {A.}~\bibnamefont {Piel}},
  \bibinfo {author} {\bibfnamefont {C.}~\bibnamefont {Schr\"oder}}, \ and\
  \bibinfo {author} {\bibfnamefont {T.}~\bibnamefont {Klinger}},\ }\href
  {\doibase 10.1103/PhysRevE.63.056401} {\bibfield  {journal} {\bibinfo
  {journal} {Phys. Rev. E}\ }\textbf {\bibinfo {volume} {63}},\ \bibinfo
  {pages} {056401} (\bibinfo {year} {2001})}\BibitemShut {NoStop}%
\bibitem [{\citenamefont {Klinger}\ \emph {et~al.}(1993)\citenamefont
  {Klinger}, \citenamefont {Piel}, \citenamefont {Seddighi},\ and\
  \citenamefont {Wilke}}]{Klinger_PLA_1993}%
  \BibitemOpen
  \bibfield  {author} {\bibinfo {author} {\bibfnamefont {T.}~\bibnamefont
  {Klinger}}, \bibinfo {author} {\bibfnamefont {A.}~\bibnamefont {Piel}},
  \bibinfo {author} {\bibfnamefont {F.}~\bibnamefont {Seddighi}}, \ and\
  \bibinfo {author} {\bibfnamefont {C.}~\bibnamefont {Wilke}},\ }\href
  {\doibase https://doi.org/10.1016/0375-9601(93)91079-K} {\bibfield  {journal}
  {\bibinfo  {journal} {Phys. Lett. A}\ }\textbf {\bibinfo {volume} {182}},\
  \bibinfo {pages} {312} (\bibinfo {year} {1993})}\BibitemShut {NoStop}%
\bibitem [{\citenamefont {Klinger}\ \emph {et~al.}(1995)\citenamefont
  {Klinger}, \citenamefont {Greiner}, \citenamefont {Rohde}, \citenamefont
  {Piel},\ and\ \citenamefont {Koepke}}]{Klinger_PRE_1995}%
  \BibitemOpen
  \bibfield  {author} {\bibinfo {author} {\bibfnamefont {T.}~\bibnamefont
  {Klinger}}, \bibinfo {author} {\bibfnamefont {F.}~\bibnamefont {Greiner}},
  \bibinfo {author} {\bibfnamefont {A.}~\bibnamefont {Rohde}}, \bibinfo
  {author} {\bibfnamefont {A.}~\bibnamefont {Piel}}, \ and\ \bibinfo {author}
  {\bibfnamefont {M.~E.}\ \bibnamefont {Koepke}},\ }\href {\doibase
  10.1103/PhysRevE.52.4316} {\bibfield  {journal} {\bibinfo  {journal} {Phys.
  Rev. E}\ }\textbf {\bibinfo {volume} {52}},\ \bibinfo {pages} {4316}
  (\bibinfo {year} {1995})}\BibitemShut {NoStop}%
\bibitem [{\citenamefont {Gyergyek}(1999)}]{Gyergyek_PPCF_1999}%
  \BibitemOpen
  \bibfield  {author} {\bibinfo {author} {\bibfnamefont {T.}~\bibnamefont
  {Gyergyek}},\ }\href@noop {} {\bibfield  {journal} {\bibinfo  {journal}
  {Plasma Phys. Controlled Fusion}\ }\textbf {\bibinfo {volume} {41}},\
  \bibinfo {pages} {175} (\bibinfo {year} {1999})}\BibitemShut {NoStop}%
\bibitem [{\citenamefont {Nurujjaman}\ and\ \citenamefont
  {Iyengar}(2010)}]{Nurujjaman_PRE_2010}%
  \BibitemOpen
  \bibfield  {author} {\bibinfo {author} {\bibfnamefont {M.}~\bibnamefont
  {Nurujjaman}}\ and\ \bibinfo {author} {\bibfnamefont {A.~N.~S.}\ \bibnamefont
  {Iyengar}},\ }\href {\doibase 10.1103/PhysRevE.82.056210} {\bibfield
  {journal} {\bibinfo  {journal} {Phys. Rev. E}\ }\textbf {\bibinfo {volume}
  {82}},\ \bibinfo {pages} {056210} (\bibinfo {year} {2010})}\BibitemShut
  {NoStop}%
\bibitem [{\citenamefont {Ruhunusiri}\ and\ \citenamefont
  {Goree}(2012)}]{Suranga_PRE_2012}%
  \BibitemOpen
  \bibfield  {author} {\bibinfo {author} {\bibfnamefont {W.~D.~S.}\
  \bibnamefont {Ruhunusiri}}\ and\ \bibinfo {author} {\bibfnamefont
  {J.}~\bibnamefont {Goree}},\ }\href {\doibase 10.1103/PhysRevE.85.046401}
  {\bibfield  {journal} {\bibinfo  {journal} {Phys. Rev. E}\ }\textbf {\bibinfo
  {volume} {85}},\ \bibinfo {pages} {046401} (\bibinfo {year}
  {2012})}\BibitemShut {NoStop}%
\bibitem [{\citenamefont {Pilch}\ \emph {et~al.}(2009)\citenamefont {Pilch},
  \citenamefont {Reichstein},\ and\ \citenamefont {Piel}}]{Pilch_POP_2009}%
  \BibitemOpen
  \bibfield  {author} {\bibinfo {author} {\bibfnamefont {I.}~\bibnamefont
  {Pilch}}, \bibinfo {author} {\bibfnamefont {T.}~\bibnamefont {Reichstein}}, \
  and\ \bibinfo {author} {\bibfnamefont {A.}~\bibnamefont {Piel}},\ }\href
  {\doibase 10.1063/1.3274928} {\bibfield  {journal} {\bibinfo  {journal}
  {Phys. Plasmas}\ }\textbf {\bibinfo {volume} {16}},\ \bibinfo {pages}
  {123709} (\bibinfo {year} {2009})}\BibitemShut {NoStop}%
\bibitem [{\citenamefont {Williams}(2014)}]{Williams_PRE_2014A}%
  \BibitemOpen
  \bibfield  {author} {\bibinfo {author} {\bibfnamefont {J.~D.}\ \bibnamefont
  {Williams}},\ }\href {\doibase 10.1103/PhysRevE.90.043103} {\bibfield
  {journal} {\bibinfo  {journal} {Phys. Rev. E}\ }\textbf {\bibinfo {volume}
  {90}},\ \bibinfo {pages} {043103} (\bibinfo {year} {2014})}\BibitemShut
  {NoStop}%
\bibitem [{\citenamefont {Chaubey}\ \emph {et~al.}(2015)\citenamefont
  {Chaubey}, \citenamefont {Mukherjee}, \citenamefont {Sekar~Iyengar},\ and\
  \citenamefont {Sen}}]{Neeraj_POP_2015}%
  \BibitemOpen
  \bibfield  {author} {\bibinfo {author} {\bibfnamefont {N.}~\bibnamefont
  {Chaubey}}, \bibinfo {author} {\bibfnamefont {S.}~\bibnamefont {Mukherjee}},
  \bibinfo {author} {\bibfnamefont {A.~N.}\ \bibnamefont {Sekar~Iyengar}}, \
  and\ \bibinfo {author} {\bibfnamefont {A.}~\bibnamefont {Sen}},\ }\href
  {\doibase 10.1063/1.4913227} {\bibfield  {journal} {\bibinfo  {journal}
  {Phys. Plasmas}\ }\textbf {\bibinfo {volume} {22}},\ \bibinfo {pages}
  {022312} (\bibinfo {year} {2015})}\BibitemShut {NoStop}%
\bibitem [{\citenamefont {Fukuyama}\ \emph {et~al.}(2006)\citenamefont
  {Fukuyama}, \citenamefont {Kozakov}, \citenamefont {Testrich},\ and\
  \citenamefont {Wilke}}]{Fukuyama_PRL_2006}%
  \BibitemOpen
  \bibfield  {author} {\bibinfo {author} {\bibfnamefont {T.}~\bibnamefont
  {Fukuyama}}, \bibinfo {author} {\bibfnamefont {R.}~\bibnamefont {Kozakov}},
  \bibinfo {author} {\bibfnamefont {H.}~\bibnamefont {Testrich}}, \ and\
  \bibinfo {author} {\bibfnamefont {C.}~\bibnamefont {Wilke}},\ }\href
  {\doibase 10.1103/PhysRevLett.96.024101} {\bibfield  {journal} {\bibinfo
  {journal} {Phys. Rev. Lett.}\ }\textbf {\bibinfo {volume} {96}},\ \bibinfo
  {pages} {024101} (\bibinfo {year} {2006})}\BibitemShut {NoStop}%
\bibitem [{\citenamefont {Chaubey}\ \emph {et~al.}(2016)\citenamefont
  {Chaubey}, \citenamefont {Mukherjee}, \citenamefont {Sen},\ and\
  \citenamefont {Iyengar}}]{Neeraj_PRE_2016}%
  \BibitemOpen
  \bibfield  {author} {\bibinfo {author} {\bibfnamefont {N.}~\bibnamefont
  {Chaubey}}, \bibinfo {author} {\bibfnamefont {S.}~\bibnamefont {Mukherjee}},
  \bibinfo {author} {\bibfnamefont {A.}~\bibnamefont {Sen}}, \ and\ \bibinfo
  {author} {\bibfnamefont {A.~N.~S.}\ \bibnamefont {Iyengar}},\ }\href
  {\doibase 10.1103/PhysRevE.94.061201} {\bibfield  {journal} {\bibinfo
  {journal} {Phys. Rev. E}\ }\textbf {\bibinfo {volume} {94}},\ \bibinfo
  {pages} {061201} (\bibinfo {year} {2016})}\BibitemShut {NoStop}%
\bibitem [{\citenamefont {Shukla}\ and\ \citenamefont
  {Eliasson}(2009)}]{Shukla_RMP_2009}%
  \BibitemOpen
  \bibfield  {author} {\bibinfo {author} {\bibfnamefont {P.~K.}\ \bibnamefont
  {Shukla}}\ and\ \bibinfo {author} {\bibfnamefont {B.}~\bibnamefont
  {Eliasson}},\ }\href {\doibase 10.1103/RevModPhys.81.25} {\bibfield
  {journal} {\bibinfo  {journal} {Rev. Mod. Phys.}\ }\textbf {\bibinfo {volume}
  {81}},\ \bibinfo {pages} {25} (\bibinfo {year} {2009})}\BibitemShut {NoStop}%
\bibitem [{\citenamefont {Morfill}\ and\ \citenamefont
  {Ivlev}(2009)}]{Morfill_RMP_2009}%
  \BibitemOpen
  \bibfield  {author} {\bibinfo {author} {\bibfnamefont {G.~E.}\ \bibnamefont
  {Morfill}}\ and\ \bibinfo {author} {\bibfnamefont {A.~V.}\ \bibnamefont
  {Ivlev}},\ }\href {\doibase 10.1103/RevModPhys.81.1353} {\bibfield  {journal}
  {\bibinfo  {journal} {Rev. Mod. Phys.}\ }\textbf {\bibinfo {volume} {81}},\
  \bibinfo {pages} {1353} (\bibinfo {year} {2009})}\BibitemShut {NoStop}%
\bibitem [{\citenamefont {Shukla}\ and\ \citenamefont
  {Mamun}(2001)}]{Shukla_IOP_2001}%
  \BibitemOpen
  \bibfield  {author} {\bibinfo {author} {\bibfnamefont {P.~K.}\ \bibnamefont
  {Shukla}}\ and\ \bibinfo {author} {\bibfnamefont {A.~A.}\ \bibnamefont
  {Mamun}},\ }\href@noop {} {\emph {\bibinfo {title} {Introduction to Dusty
  Plasma Physics}}}\ (\bibinfo  {publisher} {Institute of Physics, Bristol},\
  \bibinfo {year} {2001})\BibitemShut {NoStop}%
\bibitem [{\citenamefont {Thompson}\ \emph {et~al.}(1997)\citenamefont
  {Thompson}, \citenamefont {Barkan}, \citenamefont {D’Angelo},\ and\
  \citenamefont {Merlino}}]{Thompson_POP_1997}%
  \BibitemOpen
  \bibfield  {author} {\bibinfo {author} {\bibfnamefont {C.}~\bibnamefont
  {Thompson}}, \bibinfo {author} {\bibfnamefont {A.}~\bibnamefont {Barkan}},
  \bibinfo {author} {\bibfnamefont {N.}~\bibnamefont {D’Angelo}}, \ and\
  \bibinfo {author} {\bibfnamefont {R.~L.}\ \bibnamefont {Merlino}},\ }\href
  {\doibase 10.1063/1.872238} {\bibfield  {journal} {\bibinfo  {journal} {Phys.
  Plasmas}\ }\textbf {\bibinfo {volume} {4}},\ \bibinfo {pages} {2331}
  (\bibinfo {year} {1997})}\BibitemShut {NoStop}%
\bibitem [{\citenamefont {Schwabe}\ \emph {et~al.}(2007)\citenamefont
  {Schwabe}, \citenamefont {Rubin-Zuzic}, \citenamefont {Zhdanov},
  \citenamefont {Thomas},\ and\ \citenamefont {Morfill}}]{Schwabe_PRL_2007}%
  \BibitemOpen
  \bibfield  {author} {\bibinfo {author} {\bibfnamefont {M.}~\bibnamefont
  {Schwabe}}, \bibinfo {author} {\bibfnamefont {M.}~\bibnamefont
  {Rubin-Zuzic}}, \bibinfo {author} {\bibfnamefont {S.}~\bibnamefont
  {Zhdanov}}, \bibinfo {author} {\bibfnamefont {H.~M.}\ \bibnamefont {Thomas}},
  \ and\ \bibinfo {author} {\bibfnamefont {G.~E.}\ \bibnamefont {Morfill}},\
  }\href {\doibase 10.1103/PhysRevLett.99.095002} {\bibfield  {journal}
  {\bibinfo  {journal} {Phys. Rev. Lett.}\ }\textbf {\bibinfo {volume} {99}},\
  \bibinfo {pages} {095002} (\bibinfo {year} {2007})}\BibitemShut {NoStop}%
\bibitem [{\citenamefont {Barkan}\ \emph {et~al.}(1995)\citenamefont {Barkan},
  \citenamefont {Merlino},\ and\ \citenamefont {D’Angelo}}]{Barkan_POP_1995}%
  \BibitemOpen
  \bibfield  {author} {\bibinfo {author} {\bibfnamefont {A.}~\bibnamefont
  {Barkan}}, \bibinfo {author} {\bibfnamefont {R.~L.}\ \bibnamefont {Merlino}},
  \ and\ \bibinfo {author} {\bibfnamefont {N.}~\bibnamefont {D’Angelo}},\
  }\href {\doibase 10.1063/1.871121} {\bibfield  {journal} {\bibinfo  {journal}
  {Phys. Plasmas}\ }\textbf {\bibinfo {volume} {2}},\ \bibinfo {pages} {3563}
  (\bibinfo {year} {1995})}\BibitemShut {NoStop}%
\bibitem [{\citenamefont {Heinrich}\ \emph {et~al.}(2009)\citenamefont
  {Heinrich}, \citenamefont {Kim},\ and\ \citenamefont
  {Merlino}}]{Heinrich_PRL_2009}%
  \BibitemOpen
  \bibfield  {author} {\bibinfo {author} {\bibfnamefont {J.}~\bibnamefont
  {Heinrich}}, \bibinfo {author} {\bibfnamefont {S.-H.}\ \bibnamefont {Kim}}, \
  and\ \bibinfo {author} {\bibfnamefont {R.~L.}\ \bibnamefont {Merlino}},\
  }\href {\doibase 10.1103/PhysRevLett.103.115002} {\bibfield  {journal}
  {\bibinfo  {journal} {Phys. Rev. Lett.}\ }\textbf {\bibinfo {volume} {103}},\
  \bibinfo {pages} {115002} (\bibinfo {year} {2009})}\BibitemShut {NoStop}%
\bibitem [{\citenamefont {Bandyopadhyay}\ \emph {et~al.}(2008)\citenamefont
  {Bandyopadhyay}, \citenamefont {Prasad}, \citenamefont {Sen},\ and\
  \citenamefont {Kaw}}]{Bandyopadhyay_PRL_2008}%
  \BibitemOpen
  \bibfield  {author} {\bibinfo {author} {\bibfnamefont {P.}~\bibnamefont
  {Bandyopadhyay}}, \bibinfo {author} {\bibfnamefont {G.}~\bibnamefont
  {Prasad}}, \bibinfo {author} {\bibfnamefont {A.}~\bibnamefont {Sen}}, \ and\
  \bibinfo {author} {\bibfnamefont {P.~K.}\ \bibnamefont {Kaw}},\ }\href
  {\doibase 10.1103/PhysRevLett.101.065006} {\bibfield  {journal} {\bibinfo
  {journal} {Phys. Rev. Lett.}\ }\textbf {\bibinfo {volume} {101}},\ \bibinfo
  {pages} {065006} (\bibinfo {year} {2008})}\BibitemShut {NoStop}%
\bibitem [{\citenamefont {Deka}\ \emph {et~al.}(2017)\citenamefont {Deka},
  \citenamefont {Boruah}, \citenamefont {Sharma},\ and\ \citenamefont
  {Bailung}}]{Deka_POP_2017}%
  \BibitemOpen
  \bibfield  {author} {\bibinfo {author} {\bibfnamefont {T.}~\bibnamefont
  {Deka}}, \bibinfo {author} {\bibfnamefont {A.}~\bibnamefont {Boruah}},
  \bibinfo {author} {\bibfnamefont {S.~K.}\ \bibnamefont {Sharma}}, \ and\
  \bibinfo {author} {\bibfnamefont {H.}~\bibnamefont {Bailung}},\ }\href
  {\doibase 10.1063/1.5001721} {\bibfield  {journal} {\bibinfo  {journal}
  {Phys. Plasmas}\ }\textbf {\bibinfo {volume} {24}},\ \bibinfo {pages}
  {093706} (\bibinfo {year} {2017})}\BibitemShut {NoStop}%
\bibitem [{\citenamefont {Piel}\ and\ \citenamefont
  {Melzer}(2001)}]{Piel_PPCF_2001}%
  \BibitemOpen
  \bibfield  {author} {\bibinfo {author} {\bibfnamefont {A.}~\bibnamefont
  {Piel}}\ and\ \bibinfo {author} {\bibfnamefont {A.}~\bibnamefont {Melzer}},\
  }\href {\doibase 10.1088/0741-3335/44/1/201} {\bibfield  {journal} {\bibinfo
  {journal} {Plasma Phys. Controlled Fusion}\ }\textbf {\bibinfo {volume}
  {44}},\ \bibinfo {pages} {1} (\bibinfo {year} {2001})}\BibitemShut {NoStop}%
\bibitem [{\citenamefont {Liu}\ \emph {et~al.}(2018)\citenamefont {Liu},
  \citenamefont {Goree}, \citenamefont {Flanagan}, \citenamefont {Sen},
  \citenamefont {Tiwari}, \citenamefont {Ganguli},\ and\ \citenamefont
  {Crabtree}}]{Liu_POP_2018}%
  \BibitemOpen
  \bibfield  {author} {\bibinfo {author} {\bibfnamefont {B.}~\bibnamefont
  {Liu}}, \bibinfo {author} {\bibfnamefont {J.}~\bibnamefont {Goree}}, \bibinfo
  {author} {\bibfnamefont {T.~M.}\ \bibnamefont {Flanagan}}, \bibinfo {author}
  {\bibfnamefont {A.}~\bibnamefont {Sen}}, \bibinfo {author} {\bibfnamefont
  {S.~K.}\ \bibnamefont {Tiwari}}, \bibinfo {author} {\bibfnamefont
  {G.}~\bibnamefont {Ganguli}}, \ and\ \bibinfo {author} {\bibfnamefont
  {C.}~\bibnamefont {Crabtree}},\ }\href {\doibase 10.1063/1.5046402}
  {\bibfield  {journal} {\bibinfo  {journal} {Phys. Plasmas}\ }\textbf
  {\bibinfo {volume} {25}},\ \bibinfo {pages} {113701} (\bibinfo {year}
  {2018})}\BibitemShut {NoStop}%
\bibitem [{\citenamefont {Thompson}\ \emph {et~al.}(1999)\citenamefont
  {Thompson}, \citenamefont {Barkan}, \citenamefont {Merlino},\ and\
  \citenamefont {D'Angelo}}]{Thompson_IEEE_TPS_1999}%
  \BibitemOpen
  \bibfield  {author} {\bibinfo {author} {\bibfnamefont {C.}~\bibnamefont
  {Thompson}}, \bibinfo {author} {\bibfnamefont {A.}~\bibnamefont {Barkan}},
  \bibinfo {author} {\bibfnamefont {R.}~\bibnamefont {Merlino}}, \ and\
  \bibinfo {author} {\bibfnamefont {N.}~\bibnamefont {D'Angelo}},\ }\href@noop
  {} {\bibfield  {journal} {\bibinfo  {journal} {IEEE Trans. Plasma Sci.}\
  }\textbf {\bibinfo {volume} {27}},\ \bibinfo {pages} {146} (\bibinfo {year}
  {1999})}\BibitemShut {NoStop}%
\bibitem [{\citenamefont {Don}\ \emph {et~al.}(2014)\citenamefont {Don},
  \citenamefont {Ruhunusiri},\ and\ \citenamefont {Goree}}]{Don_IEEE_TPS_2014}%
  \BibitemOpen
  \bibfield  {author} {\bibinfo {author} {\bibfnamefont {W.}~\bibnamefont
  {Don}}, \bibinfo {author} {\bibfnamefont {S.}~\bibnamefont {Ruhunusiri}}, \
  and\ \bibinfo {author} {\bibfnamefont {J.}~\bibnamefont {Goree}},\
  }\href@noop {} {\bibfield  {journal} {\bibinfo  {journal} {IEEE Trans. Plasma
  Sci.}\ }\textbf {\bibinfo {volume} {42}},\ \bibinfo {pages} {2688} (\bibinfo
  {year} {2014})}\BibitemShut {NoStop}%
\bibitem [{\citenamefont {Feng}\ \emph {et~al.}(2007)\citenamefont {Feng},
  \citenamefont {Goree},\ and\ \citenamefont {Liu}}]{Feng_RSI_2007}%
  \BibitemOpen
  \bibfield  {author} {\bibinfo {author} {\bibfnamefont {Y.}~\bibnamefont
  {Feng}}, \bibinfo {author} {\bibfnamefont {J.}~\bibnamefont {Goree}}, \ and\
  \bibinfo {author} {\bibfnamefont {B.}~\bibnamefont {Liu}},\ }\href {\doibase
  10.1063/1.2735920} {\bibfield  {journal} {\bibinfo  {journal} {Rev. Sci.
  Instrum.}\ }\textbf {\bibinfo {volume} {78}},\ \bibinfo {pages} {053704}
  (\bibinfo {year} {2007})}\BibitemShut {NoStop}%
\bibitem [{\citenamefont {Deka}\ \emph {et~al.}(2020)\citenamefont {Deka},
  \citenamefont {Chutia}, \citenamefont {Bailung}, \citenamefont {Sharma},\
  and\ \citenamefont {Bailung}}]{Deka_PST_2020}%
  \BibitemOpen
  \bibfield  {author} {\bibinfo {author} {\bibfnamefont {T.}~\bibnamefont
  {Deka}}, \bibinfo {author} {\bibfnamefont {B.}~\bibnamefont {Chutia}},
  \bibinfo {author} {\bibfnamefont {Y.}~\bibnamefont {Bailung}}, \bibinfo
  {author} {\bibfnamefont {S.~K.}\ \bibnamefont {Sharma}}, \ and\ \bibinfo
  {author} {\bibfnamefont {H.}~\bibnamefont {Bailung}},\ }\href {\doibase
  10.1088/2058-6272/ab5b18} {\bibfield  {journal} {\bibinfo  {journal} {Plasma
  Sci. Technol.}\ }\textbf {\bibinfo {volume} {22}},\ \bibinfo {pages} {045002}
  (\bibinfo {year} {2020})}\BibitemShut {NoStop}%
\bibitem [{\citenamefont {Williams}(2018)}]{Williams_IEEE_2018}%
  \BibitemOpen
  \bibfield  {author} {\bibinfo {author} {\bibfnamefont {J.}~\bibnamefont
  {Williams}},\ }\href {\doibase 10.1109/TPS.2017.2765561} {\bibfield
  {journal} {\bibinfo  {journal} {IEEE Trans. Plasma Sci.}\ }\textbf {\bibinfo
  {volume} {46}},\ \bibinfo {pages} {806} (\bibinfo {year} {2018})}\BibitemShut
  {NoStop}%
\bibitem [{\citenamefont {Liu}\ \emph {et~al.}(2021)\citenamefont {Liu},
  \citenamefont {Goree}, \citenamefont {Schütt}, \citenamefont {Melzer},
  \citenamefont {Pustylnik}, \citenamefont {Thomas}, \citenamefont {Fortov},
  \citenamefont {Lipaev}, \citenamefont {Usachev}, \citenamefont {Petrov},
  \citenamefont {Zobnin},\ and\ \citenamefont {Thoma}}]{Liu_IEEE_2021}%
  \BibitemOpen
  \bibfield  {author} {\bibinfo {author} {\bibfnamefont {B.}~\bibnamefont
  {Liu}}, \bibinfo {author} {\bibfnamefont {J.}~\bibnamefont {Goree}}, \bibinfo
  {author} {\bibfnamefont {S.}~\bibnamefont {Schütt}}, \bibinfo {author}
  {\bibfnamefont {A.}~\bibnamefont {Melzer}}, \bibinfo {author} {\bibfnamefont
  {M.~Y.}\ \bibnamefont {Pustylnik}}, \bibinfo {author} {\bibfnamefont {H.~M.}\
  \bibnamefont {Thomas}}, \bibinfo {author} {\bibfnamefont {V.~E.}\
  \bibnamefont {Fortov}}, \bibinfo {author} {\bibfnamefont {A.~M.}\
  \bibnamefont {Lipaev}}, \bibinfo {author} {\bibfnamefont {A.~D.}\
  \bibnamefont {Usachev}}, \bibinfo {author} {\bibfnamefont {O.~F.}\
  \bibnamefont {Petrov}}, \bibinfo {author} {\bibfnamefont {A.~V.}\
  \bibnamefont {Zobnin}}, \ and\ \bibinfo {author} {\bibfnamefont {M.~H.}\
  \bibnamefont {Thoma}},\ }\href {\doibase 10.1109/TPS.2021.3123556} {\bibfield
   {journal} {\bibinfo  {journal} {IEEE Trans. Plasma Sci.}\ }\textbf {\bibinfo
  {volume} {49}},\ \bibinfo {pages} {3958} (\bibinfo {year}
  {2021})}\BibitemShut {NoStop}%
\bibitem [{\citenamefont {Van Der~Pol}(1927)}]{Van_PM_1927}%
  \BibitemOpen
  \bibfield  {author} {\bibinfo {author} {\bibfnamefont {B.}~\bibnamefont {Van
  Der~Pol}},\ }\href@noop {} {\bibfield  {journal} {\bibinfo  {journal}
  {Philos. Mag.}\ }\textbf {\bibinfo {volume} {3}},\ \bibinfo {pages} {65}
  (\bibinfo {year} {1927})}\BibitemShut {NoStop}%
\bibitem [{\citenamefont {Kovalev}\ and\ \citenamefont
  {Kovalev}(2017)}]{Kovalev_IJBC_2017}%
  \BibitemOpen
  \bibfield  {author} {\bibinfo {author} {\bibfnamefont {D.~P.}\ \bibnamefont
  {Kovalev}}\ and\ \bibinfo {author} {\bibfnamefont {P.~D.}\ \bibnamefont
  {Kovalev}},\ }\href {\doibase 10.1142/S0218127417501954} {\bibfield
  {journal} {\bibinfo  {journal} {Int. J. Bifurc. Chaos}\ }\textbf {\bibinfo
  {volume} {27}},\ \bibinfo {pages} {1750195} (\bibinfo {year}
  {2017})}\BibitemShut {NoStop}%
\bibitem [{\citenamefont {Sen}\ \emph {et~al.}(2015)\citenamefont {Sen},
  \citenamefont {Tiwari}, \citenamefont {Mishra},\ and\ \citenamefont
  {Kaw}}]{Sen_ASR_2015}%
  \BibitemOpen
  \bibfield  {author} {\bibinfo {author} {\bibfnamefont {A.}~\bibnamefont
  {Sen}}, \bibinfo {author} {\bibfnamefont {S.}~\bibnamefont {Tiwari}},
  \bibinfo {author} {\bibfnamefont {S.}~\bibnamefont {Mishra}}, \ and\ \bibinfo
  {author} {\bibfnamefont {P.}~\bibnamefont {Kaw}},\ }\href {\doibase
  https://doi.org/10.1016/j.asr.2015.03.021} {\bibfield  {journal} {\bibinfo
  {journal} {Adv. Space Res.}\ }\textbf {\bibinfo {volume} {56}},\ \bibinfo
  {pages} {429 } (\bibinfo {year} {2015})}\BibitemShut {NoStop}%
\bibitem [{\citenamefont {Kumar~Tiwari}\ and\ \citenamefont
  {Sen}(2016)}]{Sanat_POP_2016}%
  \BibitemOpen
  \bibfield  {author} {\bibinfo {author} {\bibfnamefont {S.}~\bibnamefont
  {Kumar~Tiwari}}\ and\ \bibinfo {author} {\bibfnamefont {A.}~\bibnamefont
  {Sen}},\ }\href {\doibase 10.1063/1.4941092} {\bibfield  {journal} {\bibinfo
  {journal} {Phys. Plasmas}\ }\textbf {\bibinfo {volume} {23}},\ \bibinfo
  {pages} {022301} (\bibinfo {year} {2016})}\BibitemShut {NoStop}%
\bibitem [{\citenamefont {Jaiswal}\ \emph
  {et~al.}(2016{\natexlab{a}})\citenamefont {Jaiswal}, \citenamefont
  {Bandyopadhyay},\ and\ \citenamefont {Sen}}]{Surabhi_PRE_2016}%
  \BibitemOpen
  \bibfield  {author} {\bibinfo {author} {\bibfnamefont {S.}~\bibnamefont
  {Jaiswal}}, \bibinfo {author} {\bibfnamefont {P.}~\bibnamefont
  {Bandyopadhyay}}, \ and\ \bibinfo {author} {\bibfnamefont {A.}~\bibnamefont
  {Sen}},\ }\href {\doibase 10.1103/PhysRevE.93.041201} {\bibfield  {journal}
  {\bibinfo  {journal} {Phys. Rev. E}\ }\textbf {\bibinfo {volume} {93}},\
  \bibinfo {pages} {041201} (\bibinfo {year} {2016}{\natexlab{a}})}\BibitemShut
  {NoStop}%
\bibitem [{\citenamefont {Nakamura}\ \emph {et~al.}(1999)\citenamefont
  {Nakamura}, \citenamefont {Bailung},\ and\ \citenamefont
  {Shukla}}]{Nakamura_PRL_1999}%
  \BibitemOpen
  \bibfield  {author} {\bibinfo {author} {\bibfnamefont {Y.}~\bibnamefont
  {Nakamura}}, \bibinfo {author} {\bibfnamefont {H.}~\bibnamefont {Bailung}}, \
  and\ \bibinfo {author} {\bibfnamefont {P.~K.}\ \bibnamefont {Shukla}},\
  }\href {\doibase 10.1103/PhysRevLett.83.1602} {\bibfield  {journal} {\bibinfo
   {journal} {Phys. Rev. Lett.}\ }\textbf {\bibinfo {volume} {83}},\ \bibinfo
  {pages} {1602} (\bibinfo {year} {1999})}\BibitemShut {NoStop}%
\bibitem [{\citenamefont {Nakamura}\ and\ \citenamefont
  {Sarma}(2001)}]{Nakamura_POP_2001}%
  \BibitemOpen
  \bibfield  {author} {\bibinfo {author} {\bibfnamefont {Y.}~\bibnamefont
  {Nakamura}}\ and\ \bibinfo {author} {\bibfnamefont {A.}~\bibnamefont
  {Sarma}},\ }\href {\doibase 10.1063/1.1387472} {\bibfield  {journal}
  {\bibinfo  {journal} {Phys. Plasmas}\ }\textbf {\bibinfo {volume} {8}},\
  \bibinfo {pages} {3921} (\bibinfo {year} {2001})}\BibitemShut {NoStop}%
\bibitem [{\citenamefont {Veeresha}\ \emph {et~al.}(2010)\citenamefont
  {Veeresha}, \citenamefont {Tiwari}, \citenamefont {Sen}, \citenamefont
  {Kaw},\ and\ \citenamefont {Das}}]{Veeresha_PRE_2010}%
  \BibitemOpen
  \bibfield  {author} {\bibinfo {author} {\bibfnamefont {B.~M.}\ \bibnamefont
  {Veeresha}}, \bibinfo {author} {\bibfnamefont {S.~K.}\ \bibnamefont
  {Tiwari}}, \bibinfo {author} {\bibfnamefont {A.}~\bibnamefont {Sen}},
  \bibinfo {author} {\bibfnamefont {P.~K.}\ \bibnamefont {Kaw}}, \ and\
  \bibinfo {author} {\bibfnamefont {A.}~\bibnamefont {Das}},\ }\href {\doibase
  10.1103/PhysRevE.81.036407} {\bibfield  {journal} {\bibinfo  {journal} {Phys.
  Rev. E}\ }\textbf {\bibinfo {volume} {81}},\ \bibinfo {pages} {036407}
  (\bibinfo {year} {2010})}\BibitemShut {NoStop}%
\bibitem [{\citenamefont {Jaiswal}\ \emph
  {et~al.}(2016{\natexlab{b}})\citenamefont {Jaiswal}, \citenamefont
  {Bandyopadhyay},\ and\ \citenamefont {Sen}}]{Jaiswal_POP_2016}%
  \BibitemOpen
  \bibfield  {author} {\bibinfo {author} {\bibfnamefont {S.}~\bibnamefont
  {Jaiswal}}, \bibinfo {author} {\bibfnamefont {P.}~\bibnamefont
  {Bandyopadhyay}}, \ and\ \bibinfo {author} {\bibfnamefont {A.}~\bibnamefont
  {Sen}},\ }\href {\doibase 10.1063/1.4960032} {\bibfield  {journal} {\bibinfo
  {journal} {Phys. Plasmas}\ }\textbf {\bibinfo {volume} {23}},\ \bibinfo
  {pages} {083701} (\bibinfo {year} {2016}{\natexlab{b}})}\BibitemShut
  {NoStop}%
\bibitem [{\citenamefont {Or\l{}owski}(1994)}]{Orowski_PRE_1994}%
  \BibitemOpen
  \bibfield  {author} {\bibinfo {author} {\bibfnamefont {A.}~\bibnamefont
  {Or\l{}owski}},\ }\href {\doibase 10.1103/PhysRevE.49.2465} {\bibfield
  {journal} {\bibinfo  {journal} {Phys. Rev. E}\ }\textbf {\bibinfo {volume}
  {49}},\ \bibinfo {pages} {2465} (\bibinfo {year} {1994})}\BibitemShut
  {NoStop}%
\bibitem [{\citenamefont {Rech}(2013)}]{Rech_EPJB_2013}%
  \BibitemOpen
  \bibfield  {author} {\bibinfo {author} {\bibfnamefont {P.~C.}\ \bibnamefont
  {Rech}},\ }\href@noop {} {\bibfield  {journal} {\bibinfo  {journal} {Eur.
  Phys. J. B}\ }\textbf {\bibinfo {volume} {86}},\ \bibinfo {pages} {1}
  (\bibinfo {year} {2013})}\BibitemShut {NoStop}%
\bibitem [{\citenamefont {Boyd}(2003)}]{Boyd_DP_2003}%
  \BibitemOpen
  \bibfield  {author} {\bibinfo {author} {\bibfnamefont {J.~P.}\ \bibnamefont
  {Boyd}},\ }\href@noop {} {\emph {\bibinfo {title} {Chebyshev and Fourier
  Spectral Methods}}}\ (\bibinfo  {publisher} {Dover, New York},\ \bibinfo
  {year} {2003})\BibitemShut {NoStop}%
\bibitem [{\citenamefont {Mir}\ \emph {et~al.}(2022)\citenamefont {Mir},
  \citenamefont {Tiwari},\ and\ \citenamefont {Sen}}]{Ajaz_POP_2022}%
  \BibitemOpen
  \bibfield  {author} {\bibinfo {author} {\bibfnamefont {A.}~\bibnamefont
  {Mir}}, \bibinfo {author} {\bibfnamefont {S.}~\bibnamefont {Tiwari}}, \ and\
  \bibinfo {author} {\bibfnamefont {A.}~\bibnamefont {Sen}},\ }\href {\doibase
  10.1063/5.0077638} {\bibfield  {journal} {\bibinfo  {journal} {Phys.
  Plasmas}\ }\textbf {\bibinfo {volume} {29}},\ \bibinfo {pages} {032303}
  (\bibinfo {year} {2022})}\BibitemShut {NoStop}%
\bibitem [{\citenamefont {Mir}\ \emph {et~al.}(2020)\citenamefont {Mir},
  \citenamefont {Tiwari}, \citenamefont {Goree}, \citenamefont {Sen},
  \citenamefont {Crabtree},\ and\ \citenamefont {Ganguli}}]{Ajaz_POP_2020}%
  \BibitemOpen
  \bibfield  {author} {\bibinfo {author} {\bibfnamefont {A.~A.}\ \bibnamefont
  {Mir}}, \bibinfo {author} {\bibfnamefont {S.~K.}\ \bibnamefont {Tiwari}},
  \bibinfo {author} {\bibfnamefont {J.}~\bibnamefont {Goree}}, \bibinfo
  {author} {\bibfnamefont {A.}~\bibnamefont {Sen}}, \bibinfo {author}
  {\bibfnamefont {C.}~\bibnamefont {Crabtree}}, \ and\ \bibinfo {author}
  {\bibfnamefont {G.}~\bibnamefont {Ganguli}},\ }\href {\doibase
  10.1063/5.0022482} {\bibfield  {journal} {\bibinfo  {journal} {Phys.
  Plasmas}\ }\textbf {\bibinfo {volume} {27}},\ \bibinfo {pages} {113701}
  (\bibinfo {year} {2020})}\BibitemShut {NoStop}%
\bibitem [{\citenamefont {Jensen}(2002)}]{Jensen_AJP_2002}%
  \BibitemOpen
  \bibfield  {author} {\bibinfo {author} {\bibfnamefont {R.~V.}\ \bibnamefont
  {Jensen}},\ }\href {\doibase 10.1119/1.1467909} {\bibfield  {journal}
  {\bibinfo  {journal} {Am. J. Phys.}\ }\textbf {\bibinfo {volume} {70}},\
  \bibinfo {pages} {607} (\bibinfo {year} {2002})}\BibitemShut {NoStop}%
\end{thebibliography}%
%%%%%%%%%%%%%%%%%%%%%%%%%%%%%%%%%%%%%%%%%%%%%%%%%%%%%%%%%%%%%%%%%%%%%%%%%%%%%%%%%%%%%%%%%%%%%%%%%%%%%%%%%%%%%%%%%%%%%%%%%%%%%%%%%%%%%%%%%%%%%%%%%%%%%%%%%%%%%%%%%%%%%%%%%%%%%%%%%%%%%%%%%%%%%%%%%%%%%%%%%%%%%%
\end{document}